\documentclass[aps, prl, 10pt, twocolumn, superscriptaddress, amsmath, amssymb, floatfix, longbibliography, nofootinbib, notitlepage]{revtex4-2}
\usepackage{amsbsy}
\usepackage{amsfonts}
\usepackage{amsmath}
\usepackage{amstext}
\usepackage{amsthm}
\usepackage{amssymb}
\usepackage{graphicx}
\usepackage{enumerate}
\usepackage{bm}
\usepackage{bbm}
\usepackage{braket}
\usepackage{color}
\usepackage{subcaption}
\definecolor{prlblue}{rgb}{0.18,0.19,0.57}
\usepackage{multirow}
\usepackage[colorlinks]{hyperref}
\usepackage{cleveref}
\usepackage{ifthen}
\usepackage{tikz}
\usepackage{graphicx} %
\usetikzlibrary{arrows.meta,calc,positioning,decorations.pathreplacing}
\tikzset{
  tensor/.style={draw, rounded corners=2pt, minimum size=12mm, thick, fill=gray!10},
  leg/.style={thick},
  op/.style={draw, circle, inner sep=1.2pt, thick, fill=white},
  every node/.style={font=\small}
}
 
\usetikzlibrary{positioning,calc}
\usepackage{ amsfonts, mathtools, physics,bm,url,tabularx,bbm,slashed,xcolor}
\usepackage [english]{babel}
\usepackage [autostyle, english = american]{csquotes}
\usepackage{breqn}
\usepackage{titlesec}
\usepackage{enumerate}
\usepackage{booktabs}
\MakeOuterQuote{"}
\graphicspath{{figures/}}
\newcommand{\avg}[1]{\langle#1\rangle}

\newcommand{\hfb}{\hfill \break}
\newcommand{\mc}[1]{\mathcal{#1}}
\newcommand{\floor}[1]{\left\lfloor #1 \right\rfloor}

\newcommand{\ie}{\begin{equation}\begin{aligned}}
\newcommand{\fe}{\end{aligned}\end{equation}}

\newcommand{\newsec}[1]{\textcolor{blue}{{\textit{#1} --}}}

\newtheorem*{claim*}{Claim}
\theoremstyle{remark}
\newtheorem*{remark*}{Remark}

\begin{document}

\title{An Intrinsically Gapless Topological Phase with Dipole Symmetry}
\author{Amogh Anakru}
\email{apa6106@psu.edu}
\affiliation{Department of Physics$,$ The Pennsylvania State University$,$ University Park$,$ Pennsylvania$,$ 16802$,$ USA}

\author{Linhao Li}
\email{linhaoli601@gmail.com}
\affiliation{Department of Physics$,$ The Pennsylvania State University$,$ University Park$,$ Pennsylvania$,$ 16802$,$ USA}

\author{Zhen Bi}
\email{zjb5184@psu.edu}
\affiliation{Department of Physics$,$ The Pennsylvania State University$,$ University Park$,$ Pennsylvania$,$ 16802$,$ USA}
\affiliation{Center for Theory of Emergent Quantum Matter$,$ Institute for Computational and Data Sciences$,$ The Pennsylvania State University$,$ University Park$,$ Pennsylvania$,$ 16802$,$ USA}
\begin{abstract}

We show that the boundary fractionalization familiar from gapped topological insulators can enforce bulk gaplessness in a dipole-conserving system. In a one-dimensional fermion model with charge conservation, dipole conservation, and particle-hole symmetry, we realize an intrinsically gapless symmetry-protected topological (igSPT) phase combining a dipolar Luttinger liquid with half-quantized boundary charges. This boundary structure obstructs a unique gapped bulk preserving these symmetries and lattice translation. Remarkably, the two charge-neutral edge configurations remain exactly degenerate even at finite size due to dipole conservation and particle-hole symmetry. The boundary fractionalization also manifests as shifted conformal towers in the dipole-resolved entanglement spectrum, reflecting distinct conformal boundary conditions of the Luttinger liquid. We support our analysis of the igSPT by studying exactly-solvable limits and performing DMRG computations, establishing its stability for a wide range of tuning parameters. 

\end{abstract}

\maketitle

\newsec{Introduction} Multipole symmetries extend charge conservation to higher moments of the charge density, constraining both particle motion and the possible phases of quantum matter \cite{Gromov_Multipole}. Dipole symmetry has particular physical relevance, as it naturally arises in diverse experimental contexts ranging from center-of-mass conservation in the lowest Landau level \cite{seidel2005incompressible} to strongly tilted optical lattices in cold atom setups \cite{guardado2020subdiffusion, scherg2021observing}. Such conservation laws fit into a broader framework of modulated symmetries which are closely related to fractonic matter \cite{chamon2005quantum, haah2011local, vijay2016fracton, nandkishore2019fractons, pretko2020fracton, shirley2018foliated,radicevic2018fractons, Gromov_Multipole, you2018subsystem, seiberg2020fractons,sala2020_ergodicity_from_dipole_cons,Rakovszky_Sala_SLIOM_DIPOLE,sala2022dynamics_modulated}. Beyond restricting dynamics of excitations, they also impose new constraints on the organization of quantum phases. 

Spontaneous breaking of dipole symmetry gives rise to unconventional quantum phases, including exotic superfluids \cite{LakeDBHM,PengYe1,PengYe2} and non-Fermi liquids \cite{lakeDFHM,anakru2023nonfermi}. In parallel, gapped topological phases protected by dipolar and other modulated symmetries have been constructed and classified, and related Lieb--Schultz--Mattis constraints have been investigated \cite{May-Mann_Hughes_Topo_Dipole_Cons,lake_Dipolar_SPT,ye2022lieb,oh2022rank,pace2022position,delfino2023effective,watanabe_exponential_toric_code,ho_tat_lam_dipolar_SPT_classification,ho_tat_topo_dipole_insulator_2024,yan2024generalized,10.21468/SciPostPhys.17.4.104,kim_you_han_noninvertible_holography_mspt,yao2025lattice,Ebisu:2025mtb,delfino_gauging_exponential,Li:2026aja,pace2026spacetime,bulmash2025defect,pace_bulmash_2026_LSM_Modulated,anakru2026matrix,lam-ning-2026,bi_lian_ghost_spt}. These results leave open how the topology of a gapped charge sector intertwines with the collective fluctuations of dipolar order. Previous work in 2-D revealed an interplay between the band topology of the charge sector and the Goldstone modes of spontaneously broken dipole symmetry \cite{anakru2024topological}. In 1-D, dipolar quasi-long-range order can instead support a Luttinger liquid, providing a setting to explore how charge-sector topology manifests in boundary states and constrains the gapless dipole dynamics. 

Here we show that this interplay gives rise to an intrinsically gapless symmetry-protected topological (igSPT) phase \cite{Scaffidi_gSPT,Ruben_Symmetry_Enriched_CFT,thorngren_verresen_gapless_TO,li2023intrinsically,wen2023bulk,PhysRevB.97.165114}. We construct a dipole-conserving extension of the SSH chain\cite{SSH_OG} with charge conservation and particle-hole symmetry, whose bulk forms a dipolar Luttinger liquid while its boundaries carry half-quantized charges. Integrating out the gapped charge sector generates a quantized Berry-phase term for the dipolar Luttinger liquid. This term produces a filling obstruction for the low-energy dipole degrees of freedom. The resulting emergent anomaly forbids a unique gapped bulk with the same boundary fractionalization when the internal symmetries and lattice translation are preserved. Remarkably, dipole conservation forbids hybridization between the two charge-neutral edge configurations, allowing particle-hole symmetry to protect their exact degeneracy even at finite size. The boundary fractionalization also modifies the conformal boundary conditions of the Luttinger liquid, producing shifted conformal towers in the dipole-resolved entanglement spectrum. An exactly solvable limit makes this structure explicit, while DMRG simulations establish an extended igSPT regime with neighboring trivial and symmetry-breaking phases.

\begin{figure}[h]
    \centering
    \includegraphics[width=\columnwidth]{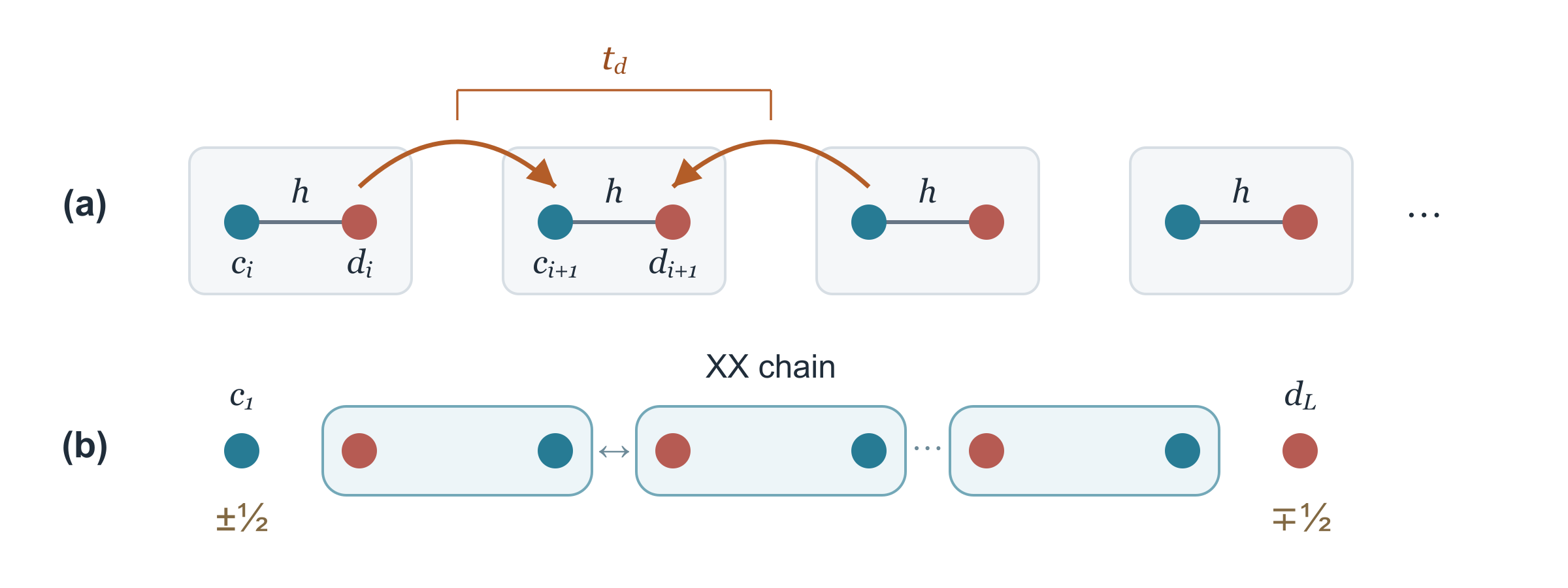}
    \caption{
    (a) Dipolar SSH model: intracell hopping $h$ and correlated dipole hopping $t_d$. Gray boxes denote unit cells. (b) In the solvable limit, singly occupied intercell pairs form an XX chain, with decoupled edge orbitals carrying charges $\pm 1/2$ and $\mp 1/2$ at neutrality.
    }
    \label{fig:dipolar_model}
\end{figure}

\newsec{Toy Model and Effective Field Theory} We consider a 1-D Hamiltonian of fermions with two orbitals $c_i,d_i$ per unit cell (see Fig. \ref{fig:dipolar_model}):
\begin{equation}\label{eq:toy_model_ssh_dipole}
\begin{aligned}
\hat{H}_{\mathrm{DSSH}}
&=
h\sum_i \left(c_i^\dagger d_i+d_i^\dagger c_i\right)
\\ &\quad-t_d\sum_{i,j}
\left(d_i^\dagger c_{i+1}\right)
A_{ij}
\left(c_{j+1}^\dagger d_j\right).
\end{aligned}
\end{equation}
where $\mathcal{A}_{ij} = \mathcal{A}_{ji}^*$. We call this the `dipolar-SSH model'.  This model has the usual $U(1)$-charge symmetry and a $U(1)$-dipole symmetry\footnote{Unless otherwise indicated, we assume an open or infinite chain, so the $U(1)$ dipole is not broken down to a discrete subgroup.} generated by 
$\hat{D} = \sum_j\,aj({n}^c_j + {n}^d_j-1)$, where $a$ is the lattice spacing. Additionally, the model has a particle-hole symmetry $\mathcal{P}^{-1}\hat{H}\mathcal{P}=\hat{H}$ where $\mathcal{P}$ maps $c_i\leftrightarrow c_i^\dag$ and $d_i\leftrightarrow -d_i^\dag$ and anticommutes with the dipole operator $\hat{D}$. The term $t_d\mc{A}_{ij}$ describes hopping for the intercell dipoles, while $h$ is an intra-cell hopping consistent with the particle-hole symmetry. Particle-hole symmetry requires $\mc{A}_{ij}$ to be real-valued, and we take $\mc{A}_{ii}=0$. We consider a filling of one fermion per unit cell and measure charge relative to a neutral background of one-half fermion per orbital. The charge of unit cell $j$ is therefore $q_j=n_{c,j}+n_{d,j}-1$, with possible values $-1,0,+1$. 

To gain some intuition for the model, we begin with a mean-field ansatz for dipolar order at large $t_d$ and subsequently include its phase fluctuations. We define the effective intercell hopping amplitude as $\Delta_i \equiv -t_d \sum_j \mathcal{A}_{ij}\avg{c_{j+1}^\dag d_j} = \abs{\Delta}e^{-i\varphi_i}$, where we neglect amplitude fluctuations. Under the $\mathcal{P}$ symmetry, the phase field transforms as $\varphi_i \rightarrow -\varphi_i$. The resulting Hamiltonian takes the form
\begin{equation}
    \label{eq:The_SSH_Model_Rotor}
    \hat{H}_{MF}
    = \sum_i \biggl(
    hc_i^\dag d_i
    + \abs{\Delta} e^{-i\varphi_i} d_i^\dag c_{i+1}
    + \mathrm{h.c.}
    \biggr)
    + \hat{H}_\varphi,
\end{equation}
where $\hat{H}_\varphi$ governs fluctuations of the compact phase field, with $\varphi \equiv \varphi+2\pi$. For a uniform, static phase configuration, the fermionic Hamiltonian is the SSH model \cite{SSH_OG}. Its topological invariant is the Berry phase $\theta = \sum_{n\in\mathrm{occ.}}\int_0^{2\pi/a} dk_x\,\mathcal{A}_n(k_x)$, with $\theta=\pi$ for $\abs{\Delta}>\abs{h}$ and $\theta=0$ for $\abs{\Delta}<\abs{h}$, modulo $2\pi$. At one fermion per unit cell, the lower band is filled and the charge sector is gapped away from $\abs{\Delta}=\abs{h}$. 

Below the charge gap, we integrate out the fermions to obtain an effective theory for $\varphi$. The phase enters the intercell hopping as a Peierls factor and therefore acts as a pseudo-gauge field $\tilde{A}_x=\varphi/a$, a correspondence used previously to describe dipolar phase fluctuations \cite{LakeDBHM,lakeDFHM,anakru2023nonfermi,anakru2024topological}. Combining the electromagnetic response of the gapped fermions with the bare dynamics in $\hat{H}_\varphi$, we obtain, to leading order in derivatives:
\begin{equation}
    \label{eq:Goldstone_LL}
    \mathcal{L}_{\mathrm{eff}}[\varphi]
    = \frac12\kappa(\partial_t\varphi)^2
    - \frac12 K(\partial_x\varphi)^2
    + \frac{\theta}{2\pi a}\partial_t\varphi.
\end{equation}
The nonuniversal coefficients $\kappa$ and $K$ describe the dipole compressibility and phase stiffness, respectively. The term linear in $\partial_t\varphi$ is a Berry-phase term inherited from the gapped charge sector, analogous to the polarization response $\frac{\theta}{2\pi}E_x$ of a one-dimensional band insulator \cite{KingSmithVanderbilt1993Polarization,VanderbiltKingSmith1993SurfaceCharge}. Its coefficient $\theta$ is the Berry phase of the occupied fermion band in Eq.~\eqref{eq:The_SSH_Model_Rotor}, quantized to $0$ or $\pi$ modulo $2\pi$ by particle-hole symmetry. The corresponding polarization is $\theta/(2\pi)$ modulo an integer, in units of the fermion charge. With this convention, the Berry term encodes a background dipole filling $-\theta/(2\pi)$ modulo one, corresponding to half-integer filling at $\theta=\pi$. Provided that symmetry-allowed phase slips are irrelevant, the resulting theory describes a dipolar Luttinger liquid with quasi-long-range order and a background dipole charge.

Since the mean-field Hamiltonian in Eq.~\eqref{eq:The_SSH_Model_Rotor} is of the SSH form, we anticipate half-quantized boundary charges in the topological regime at half-filling. These are encoded in the effective field theory in Eq.~\eqref{eq:Goldstone_LL} upon coupling to a background charge $U(1)$ gauge field. The gauging procedure is discussed in the Supplementary Material (SM). Setting $A_x=0$, this coupling amounts to replacing $\partial_t\varphi$ by $\partial_t\varphi-a\partial_xA_t$, so the topological term acquires the contribution $-\frac{\theta}{2\pi}\partial_xA_t$. We model the boundary as an interface between the topological bulk with $\theta=\pi$ and the vacuum with $\theta=0$. Integrating by parts gives the real-time boundary action
\begin{equation}
    \label{eq:Goldstone_LL_gauged_bdry}
    \mathcal{S}_{\mathrm{bdry}}
    = \frac{\Delta\theta}{2\pi}\int dt\,A_t
    = \pm\frac12\int dt\,A_t,
\end{equation}
where $\Delta\theta$ is the change in $\theta$ across the interface along the positive $x$ direction. The coefficient of $A_t$ identifies a half-integer boundary charge, defined modulo an integer. This topological contribution agrees with the polarization response of a one-dimensional band insulator \cite{qi_hughes_zhang_topological_field_theory}. In the full gauged theory, $A_x$ enters the phase-stiffness term through $\partial_x(\varphi-aA_x)$, so the response to a spatially uniform $A_x$ vanishes, consistent with dipole conservation.

\newsec{Exactly Solvable Limit}
For $h=0$ and nearest-neighbor dipole hopping, $\mathcal{A}_{ij}=\delta_{i,j+1}+\delta_{i,j-1}$, Eq.~\eqref{eq:toy_model_ssh_dipole} admits an exact mapping to an XX chain within an invariant subspace, following the construction in Ref.~\cite{moudgalya_prem_bernevig_krylov_dipole}. This subspace is defined by $n_{d,i}+n_{c,i+1}=1$ on every intercell orbital pair. Identifying $d_i^\dag\ket{0}=\ket{\downarrow}$ and $c_{i+1}^\dag\ket{0}=\ket{\uparrow}$ gives an open XX chain on $L-1$ sites, $\hat{H}=-(t_d/2)\sum_{i=1}^{L-2}(X_iX_{i+1}+Y_iY_{i+1})$, where $X_i,Y_i,Z_i$ are Pauli matrices.

The spin rotations generated by $S^z_{\mathrm{tot}}=\sum_i Z_i/2$ implement the dipole symmetry in the bulk, while particle-hole symmetry acts as spin flip, sending $S_i^z\to-S_i^z$ and $S_i^\pm\to S_i^\mp$, with $S_i^\pm=(X_i\pm iY_i)/2$. Each effective unit cell therefore carries a spin-$1/2$ projective representation of $U(1)\rtimes\mathbb{Z}_2^{\mathcal P}$. Together with translation symmetry, this gives a Lieb--Schultz--Mattis obstruction to a unique, symmetry-preserving gapped bulk ground state within the constrained subspace. At the solvable point, the bulk realizes a compact boson CFT of the form in Eq.~\eqref{eq:Goldstone_LL}, with dipolar correlations decaying as $r^{-1/2}$ and the phase-shift symmetry $\varphi\to\varphi-\alpha$ under $U_D(\alpha)=e^{i\alpha \hat{D}/a}$.

On an open chain, the mapping leaves two decoupled edge orbitals, $c_1$ and $d_L$ (Fig.~\ref{fig:dipolar_model}). At half-filling, the $L-1$ intercell pairs contain $L-1$ fermions, leaving exactly one fermion in the two edge orbitals. The resulting boundary-charge configurations, $(q_L,q_R)=(1/2,-1/2)$ and $(-1/2,1/2)$, are exactly degenerate. Symmetry-allowed perturbations can introduce bulk and boundary interactions, while nonzero $h$ generally destroys the constrained-subspace mapping. 

\newsec{Intrinsically Gapless Topology}
We now argue that the protected boundary structure is incompatible with a unique gapped bulk preserving charge, dipole, particle-hole, and translation symmetry. Opposite half-charges at the two ends of a long chain contribute a dipole moment per unit length of $1/2$ modulo an integer. In the effective theory, this is encoded by the $\theta=\pi$ term, which supplies a half-integer background filling for the dipole degree of freedom and hence an LSM filling obstruction to a symmetric gapped state \cite{lieb1961two,oshikawa2000commensurability,burnell2023filling}. In the exactly solvable limit, this obstruction is realized by a spin-$1/2$ chain with one spin per unit cell, with particle-hole symmetry enforcing zero magnetization density.

Below the charge gap, charge symmetry acts trivially on the low-energy bulk modes. The filling obstruction manifests as an emergent mixed anomaly of dipole, particle-hole, and translation symmetries\cite{thorngren_verresen_gapless_TO}. The microscopic translation--dipole algebra ties this anomaly to the topological response of the gapped charge sector (see End Matter). This response trivializes the anomaly upon restoring the full microscopic symmetry, as required for an igSPT. An independent argument in the SM rules out a translation-invariant injective MPS with these symmetries and fractionalized boundary charges.

Earlier work connected gapless topology and exact edge modes to Hilbert-space fragmentation \cite{Rakovszky_Sala_SLIOM_DIPOLE}. Here, we identify the symmetry obstruction to a gapped realization of the fractionalized boundary structure and establish an extended igSPT regime beyond the exactly solvable fragmented limit.

\begin{figure}
    \centering
    \includegraphics[width=0.9\linewidth]{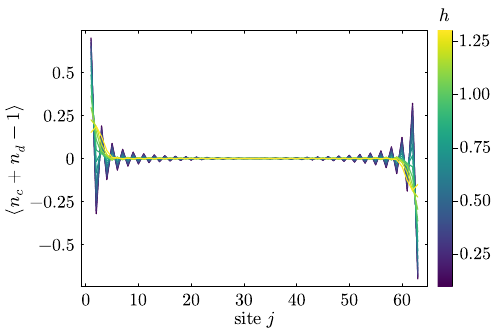}
\caption{Ground-state charge density profiles for various $h$ at $t_d=1$, $\alpha=1$, $U_0=U_1=-0.6$, and $U'=1.2$, with $L=63$ unit cells and bond dimension $\chi=256$. The boundary charge density decays exponentially into the bulk.}
\label{fig:charge_distribution}
\end{figure}

\newsec{Numerical Results on Edge Modes and Degeneracies} We use DMRG \cite{White1992DMRG,White1993DMRG,Schollwoeck2011DMRG} to study the stability of the phase under additional symmetry-allowed interactions,
\begin{equation}\label{eq:real_model_ssh_dipole}
\begin{split}
\hat{H} = \hat{H}_{\mathrm{DSSH}}
+ \sum_i \Bigl[
&U_0\tilde{n}^c_i\tilde{n}^d_i
+U_1\tilde{n}^d_i\tilde{n}^c_{i+1} \\
&+ U'\bigl(\tilde{n}^c_i\tilde{n}^c_{i+1}
+\tilde{n}^d_i\tilde{n}^d_{i+1}\bigr)
\Bigr].
\end{split}
\end{equation}
Here, $\tilde n_i^{c,d}=n_i^{c,d}-1/2$ makes particle-hole symmetry manifest. 

The ground-state charge profiles in Fig.~\ref{fig:charge_distribution} show that fractionalized boundary charges remain exponentially localized away from the exactly solvable limit. At charge neutrality, the two boundary configurations carry opposite total dipole moments and are exchanged by particle-hole symmetry. Dipole conservation forbids their hybridization, while particle-hole symmetry enforces their exact degeneracy even at finite size. In contrast, the neutral edge doublet of a conventional $U(1)\rtimes\mathbb{Z}_2$ SPT generically acquires a finite-size splitting\cite{anakru2026matrix}.

The energy cost of adding one fermion to the open chain is consistent with $\Delta E_+\sim L^{-1}$ scaling (see End Matter, Fig.~\ref{fig:edge_charge_gap}). This energy measures the finite-size splitting between edge charge sectors and is distinct from the bulk charge gap.

\begin{figure}
    \centering
    \includegraphics[width=\linewidth]{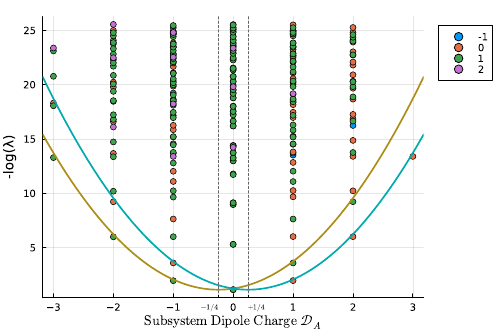}
\caption{Midpoint ($j=32$) entanglement spectrum at $h=0.8$, with other parameters as in Fig.~\ref{fig:charge_distribution}. Parabolic fits to the lowest levels in each symmetry sector use Eq.~\eqref{eq:shifted_entanglement_energies} with fixed $\delta=1/4$ and fitted $K=1.253$. Colors label the $U(1)$ charge in the subsystem.}
    \label{fig:luttinger_parabola_shifted}
\end{figure}

\newsec{Entanglement Spectrum and Conformal Boundary Conditions}
The entanglement spectrum probes both the bulk CFT \cite{lauchli_BCFT,lauchli_MPS_BCFT} and its topological boundary structure \cite{LiHaldane2008,PollmannTurnerBergOshikawa2010,TurnerPollmannBerg2011,yu2022conformal,XJYu_Entanglement_Spectrum_gSPT}. For a cut between unit cells $(j, j+1)$, we resolve it by the subsystem charge $Q_A$ and the renormalized bond dipole charge \cite{PhysRevLett.120.200602},
$\mathcal{D}_A=D_A/a-jQ_A-j\rho_0$,
where $Q_A=\sum_{i=1}^{j}q_i$, $D_A=a\sum_{i=1}^{j}i q_i$, and $q_i=n_i^c+n_i^d-1$. This observable measures the dipole moment relative to the cut after subtracting the background $\rho_0$ per unit cell. Its eigenvalues $k$ label the MPS bonds (see SM); for $\rho_0=1/2$, they are integers at even cuts and half-integers at odd cuts. The numerical spectrum exhibits two families of shifted parabolic towers,
\begin{equation}\label{eq:shifted_entanglement_energies}
E_n^\pm(k)=\frac{(k\pm\delta)^2}{2K}+n,
\qquad n\in\mathbb{Z}_{\geq0},
\end{equation}
after an overall shift and rescaling. Here, $K$ is the Luttinger parameter and $n$ labels the descendant level. Fixing $\delta=1/4$ and fitting $K$ gives good agreement with the low-lying spectrum in Fig.~\ref{fig:luttinger_parabola_shifted}, subject to finite-size and finite-bond-dimension corrections (see SM). The shift $\delta=1/4$ persists throughout the igSPT phase, up to finite-size corrections, suggesting that it reflects the topological nature of the gapless state. 

The entanglement spectrum can be obtained by boundary conformal field theory (BCFT) which relates the entanglement spectrum to the energy levels of a compact boson on an interval \cite{cardy_bcft,Cardy1986BoundaryOperatorContent}. Boundary conditions preserving the dipole $U(1)$ symmetry are Neumann for $\varphi$, or equivalently Dirichlet for its dual field $\vartheta\sim\vartheta+1$. Since the endpoint values are fixed only modulo one, the field changes across the interval by $\Delta\vartheta_s=w+\vartheta_{R,s}-\vartheta_L$, where $w\in\mathbb{Z}$ is the winding number and $s$ labels the boundary sector. The zero-mode energy is quadratic in this difference, while oscillator excitations generate the descendants, giving $E_{n,s}=(w+\vartheta_{R,s}-\vartheta_L)^2/(2K)+n$ after rescaling. The winding number $w$ and the microscopic bond label $k$ count dipole charge in the same integer steps. At an even cut, their matching gives $\delta_s\equiv\vartheta_{R,s}-\vartheta_L\pmod{1}$, as discussed in the End Matter.

The effective-spin description of the exactly solvable point gives a microscopic interpretation of the quarter shifts. On the physical edge, the fermion mode decouples, leaving a spin-flip-symmetric boundary of the spin chain. The entanglement cut splits an intercell orbital pair, producing two sectors that correspond to projections onto $S^z=\pm1/2$. In the bosonized description, the free boundary has $\vartheta_L=0$ or $1/2$ modulo one. For the polarized boundaries, the bosonization relation $S_r^z\sim(-1)^r\sin(2\pi\vartheta)$ gives $\vartheta_{R,s}=\pm1/4$ modulo one for the two opposite spin polarizations. These values give $\delta_s=\pm1/4$ modulo one and hence the quarter-shifted towers at an even cut (see End Matter).

We also expect quarter-shifted towers in an infinite-MPS description. In our iMPS boundary construction, we assume a common particle-hole-symmetric left boundary, fixing $\vartheta_L=0$ or $1/2$ modulo one. At the entanglement cut, we expect two fractionalized boundary-charge sectors exchanged by particle-hole symmetry. Half-integer dipole filling makes translation shift $\vartheta$ by $1/2$. In the minimal two-sector description, we require the two entanglement-boundary conditions to be invariant as a set under this shift, so their values differ by $1/2$. Particle-hole symmetry requires these values to have opposite signs, fixing $\vartheta_{R,s}=\pm1/4$ modulo one and yielding the quarter shifts (more details in SM). It is plausible that the entanglement spectrum of a sufficiently long open chain approaches that of the corresponding infinite MPS, as localized edge charges would have a diminishing influence on cuts deep in the bulk due to the charge gap. 

\begin{figure}
    \centering
    \includegraphics[width=0.9\linewidth]{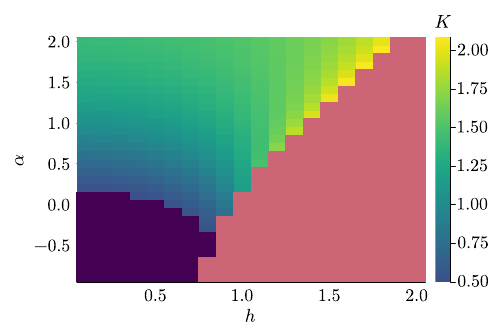}
\caption{Phase diagram at $t_d=1$, $U_0=U_1=-0.6$, and $U'=1.2$, with $L=31$, $\chi=128$. The color scale shows $K$ extracted from the mid-chain entanglement spectrum in the igSPT regime; indigo and rose-red mark the gapped CDW and trivial phases, respectively.}
    \label{fig:main_phase_diagram_K}
\end{figure}

\newsec{Global Phase Diagram}
We map out the phase diagram of Eq.~\eqref{eq:real_model_ssh_dipole} as a function of $h/t_d$ and $\alpha$, the relative strength of next-nearest-neighbor dipole hopping. We take $\mathcal{A}_{ij}=\delta_{i,j+1}+\delta_{i,j-1}+\alpha(\delta_{i,j+2}+\delta_{i,j-2})$ and fix $U_0=U_1=-0.6t_d$ and $U'=1.2t_d$. Using dipole-conserving DMRG \cite{FishmanWhiteStoudenmire2022ITensor,zaletel_mong_pollmann_dipole_dmrg}, we compare the optimized energies in the charge-neutral sectors with dimensionless total dipole quantum numbers $D=0$ and $D=-\lfloor L/2\rfloor$. Particle-hole symmetry relates the latter sector to its exactly degenerate partner at $D=+\lfloor L/2\rfloor$. Additional comparisons between dipole sectors using exact diagonalization are presented in the SM.

The resulting phase diagram, shown in Fig.~\ref{fig:main_phase_diagram_K}, reveals an extended igSPT regime with a dipolar Luttinger liquid and half-quantized boundary charges. This demonstrates that the phase persists well beyond the exactly solvable limit. We also identify a trivial gapped phase without protected boundary modes and a gapped period-$2$ charge-density-wave phase that spontaneously breaks translation symmetry. The latter is consistent with the easy-axis antiferromagnetic XXZ description in the constrained subspace at $h=\alpha=0$, where staggered pseudospin order maps to a physical charge-density wave.

The numerical results suggest a first-order transition between the igSPT and the trivial phase. Within the igSPT regime, the Luttinger parameter increases as $h$ approaches this boundary. We discuss a possible connection to proximate Lifshitz criticality in the SM, motivated by related behavior in other dipole-conserving systems \cite{lakeDFHM,zechmann2022fractonic}.

\newsec{Conclusion and Discussion} Our dipole-conserving fermion model realizes a stable igSPT phase in which half-quantized boundary charges coexist with a dipolar Luttinger liquid. The topology of the gapped charge sector produces an emergent anomaly that obstructs a symmetric gapped bulk, while dipole conservation and particle-hole symmetry protect exact finite-size edge degeneracy. The shifted entanglement towers provide a further signature of this boundary structure. 

Several questions follow from these results. Interfaces between the igSPT and trivial phases are expected to bind fractionalized charges, raising the question of how dipole conservation constrains their motion: must a moving interface emit or absorb bulk dipole excitations, and how does this affect relaxation following a quench across the topological transition \cite{fractonic_quantum_quench,zechmann2023dynamical,boesl2023deconfinement}? Although our numerics suggest a first-order transition, it remains to be understood whether additional interactions can stabilize a direct continuous transition, and how the charge-sector topology and emergent low-energy anomaly change at such a critical point. Another direction is to weakly break dipole conservation while preserving charge and particle-hole symmetries, and determine whether the igSPT evolves into a conventional gapped SPT, together with the scaling of the bulk gap and finite-size edge splitting. More broadly, these questions motivate extensions to other modulated symmetries and higher dimensions, where topology can interact with collective modes of dipolar order \cite{anakru2024topological}.

\newsec{Acknowledgements} We thank Ribhu Kaul, Ethan Lake, Hansen Wu, Sarvesh Srinivasan, Arkya Chatterjee, Sal Pace, \"Omer Aksoy, Saranesh Prembabu, Shu-Heng Shao, and Ruben Verresen for helpful discussions. In particular, AA thanks Ho Tat Lam for clarifications on dipolar SPTs and Roger Mong for elucidating the peculiarities of DMRG for dipole-conserving systems. AA and ZB acknowledge support from NSF CAREER
Grant No. DMR-2339319. LHL and ZB also acknowledge partial support from a Quantum SuperSEED grant
(ICDS QS25 029093) from the Institute for Computational and Data Sciences (ICDS) at the Pennsylvania State University. The authors recognize the Penn State ICDS  (RRID:SCR\_025154) for providing access to computational research infrastructure (RRID:SCR\_026424). We thank OpenAI for providing access through the ChatGPT for Academic Researchers program and acknowledge the use of its tools for scientific discussions and manuscript editing. 

\bibliographystyle{unsrt} 
\bibliography{mspt_reupload}%

\section{End Matter}
\newsec{Topological Response Theory}
We explain how the topology of the gapped charge sector requires an anomalous low-energy sector, following Ref.~\cite{thorngren_verresen_gapless_TO}. We use the cochain response of the occupation-number tensor-product description to capture the projective-symmetry obstruction; a complete fermionic partition function would additionally require a spin structure.

Let $A_C,A_D,A_P$ denote background one-cochains for the charge $\mathbb{Z}_2$ subgroup, the dipole $\mathbb{Z}_2$ subgroup, and particle-hole symmetry, respectively. The dipole rotation is normalized using the dipole moment in units of $a$. The translation background $A_{Tr}$ is integer-valued, with reduction modulo two understood in cup products. The half-quantized boundary charge corresponds to the charge-sector response
\begin{equation}\label{eq:TFT_MAIN}
S_{\mathrm{top}}=i\pi\int_{X_2}A_C\cup A_P,
\end{equation}
whose boundary interpretation we justify below. With only charge and particle-hole backgrounds present, $A_C\cup A_P$ is closed and defines an independent SPT response.

Dipole symmetry and translation modify this conclusion. Translating a dipole $\pi$ rotation multiplies it by a charge $\pi$ rotation, so their background fields obey
\begin{equation}\label{eq:dipole_background_constraint}
\delta A_C=A_D\cup A_{Tr}\pmod 2,
\end{equation}
while $A_D$, $A_P$, and $A_{Tr}$ are flat. Consequently,
\begin{equation}\label{eq:TFT_MANIP}
\begin{split}
\delta(A_C\cup A_P)
&=(\delta A_C)\cup A_P-A_C\cup\delta A_P\\
&=A_D\cup A_{Tr}\cup A_P
\pmod 2.
\end{split}
\end{equation}
The charge-sector term is therefore no longer a standalone SPT response for the full symmetry group \cite{SciPostPhys.17.1.013}. Its failure to be closed is precisely the mixed anomaly represented by
\begin{equation}\label{eq:TFT_LSM}
S_{\mathrm{inflow}}
=i\pi\int_{\mathcal M_3}A_D\cup A_{Tr}\cup A_P,
\qquad \partial\mathcal M_3=X_2.
\end{equation}
This anomaly is nontrivial for the low-energy symmetry: unit holonomies of the three backgrounds along independent cycles of a three-torus give an inflow phase of $-1$.

Below the bulk charge gap, charge symmetry acts trivially on the gapless bulk modes. Their remaining symmetries thus carry an anomaly that cannot be canceled using only low-energy background fields. Restoring the charge background supplies the required counterterm: Eq.~\eqref{eq:TFT_MANIP} makes the anomaly a coboundary of the full microscopic symmetry, allowing the charge-sector response to cancel the anomalous gauge variation of the low-energy theory. There is no contradiction, because not every low-energy background can be lifted to satisfy Eq.~\eqref{eq:dipole_background_constraint}. This is the anomaly-vanishing mechanism of Ref.~\cite{thorngren_verresen_gapless_TO}. The gapped charge sector leaves behind a topological response essential to the consistency of the combined theory.

Finally, the boundary representation fixes the nontrivial coefficient in Eq.~\eqref{eq:TFT_MAIN}. Particle-hole symmetry exchanges the boundary states with charges $q=\pm1/2$, so the boundary charge rotation $V_C=e^{i\pi q}$ and particle-hole operation $V_P$ satisfy
\begin{equation}
V_PV_CV_P^{-1}=V_C^{-1}=-V_C.
\end{equation}
This projective anticommutation is encoded by $A_C\cup A_P$. Thus, the half-quantized boundary charges select the nontrivial response, while the translation--dipole algebra ties it to the low-energy anomaly. Retaining this topology and all the protecting symmetries excludes a unique gapped bulk, consistently with the independent MPS obstruction in the SM.

\begin{figure}
    \centering
    \includegraphics[width=\linewidth]{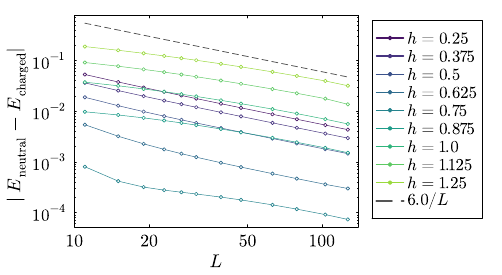}
    \caption{Edge charge gap from DMRG as a function of $L$; all other parameters are as in Fig. \ref{fig:charge_distribution}.}
    \label{fig:edge_charge_gap}
\end{figure}

\newsec{Edge Splitting}
We examine the finite-size energies associated with the four boundary-charge configurations $(q_L,q_R)=(\pm1/2,\pm1/2)$. At charge neutrality, the configurations $(+1/2,-1/2)$ and $(-1/2,+1/2)$ form an exactly degenerate doublet: dipole conservation forbids their hybridization, while particle-hole symmetry exchanges them and enforces equal energies. The configurations $(+1/2,+1/2)$ and $(-1/2,-1/2)$ form a second exactly degenerate doublet in the total-charge sectors $Q=+1$ and $Q=-1$, respectively. Here, charge conservation forbids mixing and particle-hole symmetry again ensures degeneracy. Consequently, the energies required to add and remove one fermion are equal, $\Delta E_-=\Delta E_+$. Figure~\ref{fig:edge_charge_gap} shows that the separation between the two doublets approaches $\Delta E_+\propto L^{-1}$ for the generic parameter values studied.

The observed $1/L$ splitting is consistent with the conformal finite-size spectrum of the dipolar Luttinger liquid: different boundary-charge configurations select boundary sectors whose lowest energies generically differ by $O(L^{-1})$ \cite{cardy_bcft,Ruben_Symmetry_Enriched_CFT}. The prefactor depends on microscopic parameters and can vanish at special decoupled limits. Thus, the two edge doublets become degenerate in the thermodynamic limit. 

As $h$ approaches the igSPT--trivial phase boundary, the asymptotic $1/L$ behavior emerges only at progressively larger system sizes, indicating a growing crossover length. This is compatible with a nearby critical regime but does not by itself establish a proximate continuous transition or determine its universality class. We discuss a possible connection to Lifshitz criticality in the SM.

\newsec{Bond Dipole Charge and Winding Number} We relate the microscopic bond dipole charge to the compact-boson winding number. For $A=\{1,\ldots,j\}$, define $Q_A=\sum_{i=1}^{j}q_i$ and $D_A=a\sum_{i=1}^{j}i q_i$, where $q_i=n_i^c+n_i^d-1$. Measuring the dipole moment relative to the cut and subtracting the background gives
\begin{equation}
\mathcal D_A=\frac{D_A}{a}-jQ_A-j\rho_0.
\label{eq:renormalized_bond_dipole}
\end{equation}
For $\rho_0=1/2$, its eigenvalues satisfy $k\in\mathbb Z+\epsilon_j$, with $\epsilon_j=0$ for even cuts and $1/2$ for odd cuts. An interior dipole operator $B_r=c_{r+1}^{\dagger}d_r$, with $1\leq r<j$, obeys $[\mathcal D_A,B_r]=B_r$, so increasing $k$ by one adds one unit of dipole charge.

In the Luttinger liquid, dipole rotations act as $\varphi\to\varphi-\alpha$, making $-\Pi_\varphi$ the dipole charge density. With $\partial_x\vartheta=-(\Pi_\varphi-\bar\Pi_\varphi)$, the excess bulk dipole charge is
\begin{equation}
Q_{\mathrm{LL}}^{\mathrm{ex}}=\int_{x_L}^{x_R}dx\,\partial_x\vartheta=w+\vartheta_R-\vartheta_L,\qquad w\in\mathbb Z.
\end{equation}
Since $\vartheta\sim\vartheta+1$, the Dirichlet endpoints determine the field difference only modulo one; $w$ counts the additional integer winding. The zero-mode energy is quadratic in this difference, while oscillator excitations supply the descendants, giving $E_n=(w+\vartheta_R-\vartheta_L)^2/(2K)+n$ after shifting and rescaling.

The bond label $k\in\mathbb{Z}+\epsilon_j$ and winding number $w\in\mathbb{Z}$ count dipole charge in the same unit steps, but their origins may differ:
\begin{equation}
w=k-\epsilon_j+m_s,\qquad m_s\in\mathbb{Z},
\end{equation}
where $s$ labels the boundary sector. The BCFT spectrum then becomes
\begin{equation}
E_{n,s}(k)=\frac{(k+\delta_s)^2}{2K}+n,\qquad \delta_s=\vartheta_{R,s}-\vartheta_L-\epsilon_j+m_s,
\label{eq:direct_cut_ES}
\end{equation}
with $n\in\mathbb{Z}_{\geq0}$. At an even cut, $\epsilon_j=0$, so $\delta_s\equiv\vartheta_{R,s}-\vartheta_L\pmod{1}$. The integers $m_s$ are fixed by matching the microscopic bond labels. We next identify the boundary conditions at the exactly solvable point.

\newsec{Microscopic Entanglement Boundary}
At the nearest-neighbor solvable point, $h=U_0=U_1=U'=0$, each intercell pair satisfying $n_i^d+n_{i+1}^c=1$ defines an effective spin through $|\uparrow\rangle_i=c_{i+1}^{\dagger}|0\rangle$ and $|\downarrow\rangle_i=d_i^{\dagger}|0\rangle$, with $S_i^z=(n_{i+1}^c-n_i^d)/2$. The bulk is an open XX chain of $L-1$ spins, while the edge orbitals $c_1$ and $d_L$ decouple. For odd $L$, we choose the spin-flip-symmetric ground state at zero magnetization. Fixing a charge-neutral physical-edge configuration breaks particle-hole symmetry of the full fermionic state but leaves this effective-spin ground state unchanged.

The subsystem $A=\{1,\ldots,j\}$ contains $c_1$, $j-1$ complete effective spins, and $d_j$. The cut splits the $j$-th effective spin as
\begin{equation}
|s\rangle_j
\longmapsto
\left|\frac12-s\right\rangle_{d_j}
\otimes
\left|\frac12+s\right\rangle_{c_{j+1}},
\qquad s=\pm\frac12,
\label{eq:direct_cut_embedding}
\end{equation}
where $s$ is the eigenvalue of $S_j^z$. Tracing out $c_{j+1}$ removes coherences between the two occupations. Suppressing the pure-state factor from $c_1$, which does not affect the Schmidt eigenvalues, gives
\begin{equation}
\rho_A
=\frac12\sum_{s=\pm1/2}
\rho_{j-1}^{(s)}
\otimes
\left|\frac12-s\right\rangle_{d_j}
\left\langle\frac12-s\right|,
\label{eq:direct_cut_blocks}
\end{equation}
where $\rho_{j-1}^{(s)}
=2\,{}_j\langle s|\rho^{\mathrm{spin}}_{1,\ldots,j}|s\rangle_j$
is the normalized conditional density matrix of the complete effective spins. Spin flip exchanges the two blocks and ensures their equal weights.

We interpret these blocks using boundary conformal field theory (BCFT). In the scaling limit, the XX chain is described by dual compact fields $\varphi\sim\varphi+2\pi$ and $\vartheta\sim\vartheta+1$, with bosonization relations
\begin{equation}
\begin{aligned}
S_r^x+iS_r^y&\sim e^{-i\varphi(x)},\\
S^z_{r,\mathrm{stag}}&\sim(-1)^r\sin(2\pi\vartheta(x)),\\
(-1)^r\vec S_r\cdot\vec S_{r+1}
&\sim\cos(2\pi\vartheta(x)).
\end{aligned}
\label{eq:CFT}
\end{equation}
The $U(1)$ spin rotation shifts $\varphi\to\varphi-\alpha$, while spin flip sends $\vartheta\to-\vartheta$. The $U(1)$-preserving conformal boundary conditions are Neumann for $\varphi$, $\partial_x\varphi=0$, or equivalently Dirichlet for $\vartheta$, specified by $\vartheta_L$ and $\vartheta_{R,s}$.

The free left boundary is common to both sectors and preserves spin flip. Its value satisfies $\vartheta_L=-\vartheta_L$ modulo one, giving
\begin{equation}
\vartheta_L=0\ \text{or}\ \frac12
\qquad (\mathrm{mod}\ 1).
\end{equation}
At the cut, projection onto $S_j^z=s$ selects a definite spin polarization. In the polarized-boundary description, the staggered-spin correspondence in Eq.~\eqref{eq:CFT} assigns
\begin{equation}
\vartheta_{R,s}=(-1)^j\frac{s}{2}
\qquad (\mathrm{mod}\ 1).
\label{eq:direct_cut_theta}
\end{equation}
The two sectors therefore correspond to opposite Dirichlet boundary values $\vartheta_{R,s}=\pm1/4$ modulo one. 

Combining these boundary values with the winding--charge correspondence gives
\begin{equation}
\delta_s\equiv\vartheta_{R,s}-\vartheta_L-\epsilon_j
\equiv\pm\frac14\pmod{1}.
\end{equation}
Thus, this boundary identification determines the quarter-integer fractional offsets in Eq.~\eqref{eq:direct_cut_ES}.

\clearpage
\onecolumngrid

\begin{center}
\textbf{Supplemental Material}
\end{center}

\setcounter{section}{0}
\renewcommand{\thesection}{\Alph{section}}
\renewcommand{\thesubsection}{\Alph{section}\arabic{subsection}}

\setcounter{secnumdepth}{2}
\renewcommand{\thesection}{\arabic{section}}
\renewcommand{\thesubsection}{\thesection.\arabic{subsection}}
\setcounter{section}{0}

\section{Mean-Field Derivation of Luttinger Liquid Action}

Here, we show how to derive the Luttinger liquid action for the dipolar Goldstone boson, as well as the background density term. We start with the microscopic model:

\begin{equation}\label{eq:App_Micro_Model_SSH_UNIFORM}
    \hat H=\sum_i h(c_i^\dag d_i+d_i^\dag c_i)
    -t_d\sum_{ij}(d_i^\dag c_{i+1})\mathcal A_{ij}(c_{j+1}^\dag d_j)+\hat H_U.
\end{equation}
where $\hat{H}_{U}$ represents any Hubbard-like interaction terms. We notice that the operators $d_i^\dag c_{i+1}$ transform by a phase $e^{-i\beta}$ under the $U(1)$ dipole symmetry $\exp(i\beta\hat D/a)$, so they are dipole creation/annihilation operators\footnote{We refer to them as such due to their symmetry charges, but we caution that these do not obey canonical commutation relations.}, and $\mathcal{A}_{ij}$ describes the hopping of these dipoles. Anticipating some sort of dipole order, we perform a Hubbard-Stratonovich transform to decouple the four-fermion dipole-hopping term as follows:

\begin{equation}\label{eq:App_Micro_Model_SSH_UNIFORM_ACTION}
\begin{split}
S=\int dt\biggl\{&\sum_i\Bigl[i\Psi_i^\dag\partial_t\Psi_i
-h(c_i^\dag d_i+d_i^\dag c_i)
-\bigl(\Delta_i d_i^\dag c_{i+1}+\mathrm{h.c.}\bigr)\Bigr]\\
&-\frac{1}{t_d}\sum_{ij}\Delta_i^*[\mathcal A^{-1}]_{ij}\Delta_j-H_U\biggr\}.
\end{split}
\end{equation}
with $\Delta_i=-t_d\sum_j\mathcal A_{ij}\avg{c_{j+1}^\dag d_j}$.
The Hubbard--Stratonovich identity is understood as an oscillatory
real-time Gaussian integral with a convergence prescription in each
nonzero eigenmode of $\mathcal A$. If zero eigenvalues are present,
$\mathcal A^{-1}$ denotes the inverse restricted to the nonzero-eigenvalue
subspace, and no auxiliary field is introduced in the zero modes.
The following long-wavelength expansion concerns the uniform channel
and does not assume that $\mathcal A$ is positive definite at all momenta.
Now, we suppose that $\mathcal{A}_{ij}$ is such that the Fourier transform $\frac{1}{t_d}\sum_{j}[\mathcal{A}^{-1}]_{kj} e^{i(j-k)q}\sim r_0 + \frac12Mq^2$ with $r_0, M>0$. In order to perform a mean-field treatment, we consider uniform fluctuations of the order parameter field $\Delta$ and construct an effective action for these fluctuations. If we neglect the Hubbard interaction $H_U$, one can show that the effective energy functional to be minimized is

\begin{equation}\label{eq:uniform_energy_functional}
E(\abs{\Delta}) = r_0\abs{\Delta}^2 + E_0(\abs{\Delta})
\end{equation}
where $E_0$ is the energy of the half-filled free-fermion ground state defined by the Bloch Hamiltonian $\sum_i(hc_i^\dag d_i+\Delta d_i^\dag c_{i+1}+\mathrm{h.c.})$. This energy is given by an elliptic integral, and we compute

\begin{equation}\label{eq:uniform_energy_functional_expanded}
E(\abs{\Delta}) = \biggr(r_0-\frac{1}{4h}\biggr)\abs{\Delta}^2 -\frac{1}{64h^3}\abs{\Delta}^4 + \ldots
\end{equation}
Recall that $r_0\sim 1/t_d$. Intuitively, when $h\gg t_d$, we have $r_0\gg \frac{1}{4h}$ and the solution with $\abs{\Delta}=0$ is a stable minimum. As $t_d$ grows, this minimum becomes unstable allowing for a solution with $\Delta\neq 0$. Thus, dipole symmetry ordering is driven by a competition between the on-site term $h$ and the dipole-hopping term $t_d$. Because of the sign of the $\abs{\Delta}^4$ term, we expect that the ordering transition is first-order.

Supposing we obtain a solution $\Delta\neq 0$, then to capture the effects of the most relevant (Goldstone) fluctuations, we replace $\Delta_i\sim |\Delta|e^{-i\varphi_i}$, and obtain
\begin{equation}\label{eq:Mid_IR_ungauged}
\begin{split}
S_{eff}=\int dt\biggl\{&\sum_i\Bigl[i\Psi_i^\dag\partial_t\Psi_i
-h(c_i^\dag d_i+d_i^\dag c_i)
-|\Delta|\bigl(e^{-i\varphi_i}d_i^\dag c_{i+1}+\mathrm{h.c.}\bigr)\Bigr]\\
&-\frac{|\Delta|^2}{t_d}\sum_{ij}e^{i\varphi_i}[\mathcal A^{-1}]_{ij}e^{-i\varphi_j}\biggr\}.
\end{split}
\end{equation}
Now, we recognize that the bottom term represents a gradient energy for the Goldstone boson. Using translation invariance and Hermiticity of $\mathcal{A}_{ij}$, we can expand in powers of $(\varphi_i - \varphi_j)$. In the continuum limit, this corresponds to a term like $\frac{K_0}{2}(\partial_x\varphi)^2$. 

Now supposing that $|\Delta|\neq h$, we can also integrate out the gapped fermions and construct an effective action for $\varphi$ order-by-order. The quadratic part of this action can be obtained by summing up finitely many diagrams for the 2-point $\varphi$ correlator. Upon recognizing that $\varphi$ couples to the fermions as a Peierls phase, we can write the quadratic effective action 

\begin{equation}\label{eq:Quadratic_EFT_ungauged}
\begin{split}
S_{eff}[\varphi] &= \int dtdx\biggr(-\frac12 K_0(\partial_x\varphi)^2 + \delta\mathcal{L}_{EM}[A^{(\varphi)}_\mu]\biggr) 
\end{split}
\end{equation}
where $\delta\mathcal{L}_{EM}[A_\mu]$ is the contribution to the effective electrodynamic action obtained from integrating out all matter that couples to $A_\mu$, and $A^{(\varphi)}_\mu = \delta^x_\mu\varphi/a$ (here $a$ is a lattice constant). The computation of the leading terms in $\delta\mathcal{L}_{EM}$ for free fermions is a textbook result \cite{qi_hughes_zhang_topological_field_theory,Bernevig_Book_2013} and we may simply state the result upon recognizing that the first line of Eq. \ref{eq:Mid_IR_ungauged} is the free-fermion SSH model \cite{SSH_OG}:

\begin{equation}\label{eq:Quadratic_EFT_ungauged_Final}
\begin{split}
S_{eff}[\varphi] &= \int dtdx\biggr(\frac{\kappa}{2}(\partial_t\varphi)^2-\frac12 K_0(\partial_x\varphi)^2 + \frac{\theta}{2\pi a}\partial_t\varphi\biggr)
\end{split}
\end{equation}
The $\kappa$ term is proportional to the dipole compressibility and is non-universal. The `$\theta$-term' (with $\partial_t\varphi$) is precisely the Berry phase $\theta$ for the fully-occupied band around the Brillouin zone; the corresponding polarization of the 1-D free fermions is $\theta/(2\pi)$ modulo one. With the phase convention $\Delta_i=|\Delta|e^{-i\varphi_i}$, the background dipole filling is $-\theta/(2\pi)$ modulo one. The $U(1)\rtimes\mathbb{Z}_2$ symmetry of the SSH model forces $\theta=0,\pi\,\,\mathrm{mod}\,\,2\pi$. (In fact, it is not necessary to appeal to free fermion physics to constrain $\theta$. We see from Eq. \ref{eq:Mid_IR_ungauged} that the $\mathbb{Z}_2$ particle-hole symmetry also flips $\varphi\to-\varphi$, thus imposing this symmetry on the Luttinger liquid action alone will lead to the same restriction on $\theta$.)

\subsection{Coupling to Gauge Fields}

In the main text, we discussed how the $U(1)$ gauge field couples to the dipolar Goldstone theory and used this to argue for the existence of edge modes. Here, we discuss how this coupling arises. 

Supposing that the fermion operators $c,d$ are as defined previously, we can define a time-dependent gauge transformation

\begin{equation}\label{eq:Gauge_Transform_Def}
\begin{pmatrix}c_j \\ d_j\end{pmatrix}=\Psi_j \to \Psi_j e^{i\lambda_j(t)} 
\end{equation}
for $j\in\mathbb{Z}$. Suppose we have an ordinary $U(1)$ gauge field $A_t,A_x$; the $A_t$ will live on lattice sites $j$ and the $A_x$ will live on the links $j+\frac12$. They will transform according to $A_t(j)\to A_t(j)+\partial_t\lambda_j$ and $A_x(j+\frac12)\to A_x(j+\frac12) + \lambda_{j+1}-\lambda_j$. Accordingly, the gauged model is

\begin{equation}\label{eq:App_Micro_Model_SSH_UNIFORM_GAUGED}
\begin{split}
\hat H[A_t,A_x]={}&\sum_i h(c_i^\dag d_i+d_i^\dag c_i)
-\sum_i(n_i^c+n_i^d)A_t(i)\\
&-t_d\sum_{ij}(d_i^\dag e^{-iA_x(i+\frac12)}c_{i+1})\mathcal A_{ij}
(c_{j+1}^\dag e^{iA_x(j+\frac12)}d_j).
\end{split}
\end{equation}
where we neglect any possible Hubbard interactions for brevity, as they do not affect the following discussion. Note that the subgroup of gauge transformations with $\lambda_j = \lambda_0 + \lambda_1 j$ are genuine symmetries (charge and dipole) of our ungauged Hamiltonian; accordingly, any would-be dipolar Goldstone field $\varphi$ transforms as $\varphi_{j+\frac12}\to \varphi_{j+\frac12}+\lambda_{j+1}-\lambda_j$ as well. This implies that in a low-energy field theory, $\varphi$ couples to the fermion fields in the same way as $A_x$, while $A_x$ only couples to the theory through higher-derivative couplings to $\varphi$. Thus our dipole-conserving model has a vanishing zero-momentum $U(1)$ current response, which is a general property of dipole-conserving systems\footnote{Following \cite{LakeDBHM}, write
$\hat D^a=\int d^dx\,x_a\rho(x)$ and use
$\partial_t\rho+\nabla\cdot\mathbf j=0$.
When the boundary terms vanish, dipole conservation gives
\[
\begin{aligned}
0=\partial_t\hat D^a
&=i[\hat H,\hat D^a]\\
&=-\int d^dx\,x_a\nabla\cdot\mathbf j
=\int d^dx\,j^a.
\end{aligned}
\]
Thus the spatially integrated charge current vanishes.}. We can explicitly derive the coupling of the low-energy field theory to gauge fields by performing a Hubbard-Stratonovich transformation as in the previous section:

\begin{equation}\label{eq:App_Micro_Model_SSH_UNIFORM_GAUGED_ACTION}
\begin{split}
S=\int dt\biggl\{&\sum_i\Bigl[i\Psi_i^\dag\partial_t\Psi_i+(n_i^c+n_i^d)A_t(i)
-h(c_i^\dag d_i+d_i^\dag c_i)
-\bigl(\Delta_i d_i^\dag c_{i+1}+\mathrm{h.c.}\bigr)\Bigr]\\
&-\frac{1}{t_d}\sum_{ij}\Delta_i^*[\mathcal M^{-1}]_{ij}\Delta_j\biggr\},\qquad
\mathcal M_{ij}=e^{-iA_x(i+\frac12)}\mathcal A_{ij}e^{iA_x(j+\frac12)}.
\end{split}
\end{equation}

with $\Delta_i=-t_d\sum_j\mathcal M_{ij}\avg{c_{j+1}^\dag d_j}$. At mean-field level, we replace $\Delta_i\sim |\Delta|e^{-i\varphi_i}$, and obtain
\begin{equation}\label{eq:Mid_IR_Gauged}
\begin{split}
S_{eff}=\int dt\biggl\{&\sum_i\Bigl[i\Psi_i^\dag(\partial_t-iA_t(i))\Psi_i
-h(c_i^\dag d_i+d_i^\dag c_i)
-|\Delta|\bigl(e^{-i\varphi_i}d_i^\dag c_{i+1}+\mathrm{h.c.}\bigr)\Bigr]\\
&-\frac{|\Delta|^2}{t_d}\sum_{ij}e^{i(\varphi_i-A_x(i+\frac12))}
[\mathcal A^{-1}]_{ij}e^{-i(\varphi_j-A_x(j+\frac12))}\biggr\}.
\end{split}
\end{equation}

Writing the lattice link field as $A_x(i+\frac12)\simeq aA_x(x)$ in terms of the continuum gauge field, we can see that $A_x$ only couples to the eventual low-energy field theory via the bare kinetic term for the dipole Goldstone, i.e. the $K_0(\partial_x\varphi)^2$ term transforms to $K_0(\partial_x(\varphi-aA_x))^2$. The first line of the action is the only part of the action containing fermion fields, and it takes the form of free fermions coupled to a $U(1)$ gauge field configuration $A_\mu = (A_t,\varphi/a)$. Accordingly, upon integrating out the fermions, a textbook result yields the action from the main text:

\begin{equation}\label{eq:Goldstone_LL_gauged_APP}
\mathcal{L}_{eff}[\varphi] = \frac12\kappa(\partial_t\varphi-a\partial_xA_t)^2 - \frac12 K_0(\partial_x\varphi-a\partial_xA_x)^2 + \frac{\theta}{2\pi a}(\partial_t\varphi-a\partial_xA_t).
\end{equation}

\subsection{Response Functions}

We set $a=1$ in this subsection and compute equal-time structure
factors from Euclidean correlation functions at zero background gauge
field. With our phase convention, the dipole-density fluctuation is
$\delta\rho_d=-(\Pi_\varphi-\bar\Pi_\varphi)$, where
$\Pi_\varphi$ is the real-time canonical momentum. Its Euclidean
correlator follows from the phase-space action
\[
S_E=\int d\tau dx\left[
 i\delta\rho_d\partial_\tau\varphi
 +\frac{(\delta\rho_d)^2}{2\kappa}
 +\frac{K_0}{2}(\partial_x\varphi)^2\right],
\]
where the background Berry term does not affect the Gaussian
nonzero-momentum correlators. Integrating the canonical density gives
\begin{equation}\label{eq:dipole_compressibility_1}
\begin{split}
D_E(p,\omega_E)
&=\bigl\langle\delta\rho_d(-p,-\omega_E)
\delta\rho_d(p,\omega_E)\bigr\rangle_E\\
&=\kappa-\frac{\kappa^2\omega_E^2}
{\kappa\omega_E^2+K_0p^2}
=\frac{K_0p^2}{\omega_E^2+v^2p^2},
\qquad v^2=\frac{K_0}{\kappa}.
\end{split}
\end{equation}
The contact term $\kappa$ ensures $D_E(0,\omega_E)=0$ for
$\omega_E\neq0$, as required by dipole conservation. The equal-time
dipole structure factor is
\begin{equation}\label{eq:dipole_structure_factor}
S_d(p)=\int_{-\infty}^{\infty}\frac{d\omega_E}{2\pi}
D_E(p,\omega_E)
=\frac12\sqrt{\kappa K_0}|p|
=\frac{K}{2\pi}|p|.
\end{equation}
Here $K=\pi\sqrt{\kappa K_0}$ is the Luttinger parameter, with
$\Delta(e^{i\varphi})=1/(4K)$. Using the low-energy relation
$\delta\rho=-\partial_x\delta\rho_d$, the equal-time charge
structure factor is
\begin{equation}\label{eq:charge_structure_factor}
S_c(p)=p^2S_d(p)=\frac{K}{2\pi}|p|^3.
\end{equation}
In practice, the charge structure factor is slightly easier to measure
than the dipole structure factor.

\section{Obstruction to an Injective MPS}

In the main text, we illustrated the obstruction to realizing fractionalized edge charges (i.e. those in the SSH model) in the presence of dipole symmetry. The argument there was that the edge charges contribute a half-quantized dipole density, which results in an LSM obstruction to trivially gapping out the bulk. Here, we offer a different perspective in terms of matrix product states (MPS). 

First, let us briefly review MPS. We point the reader to Refs.~\cite{perez2006matrix, pedagogical_pushthrough,cirac_review} for a complete review, and below we will also review the arguments in \cite{ho_tat_lam_dipolar_SPT_classification} that allow us to talk about MPS with dipole symmetries.

A translationally invariant matrix product state is generated by the three-legged tensor $A^s$ depicted in Eq.~\eqref{eq:MPS_intro}; here the index $s$ describes the vertical (physical) leg and $A^s$ is then a matrix in bond space (indicated by horizontal legs). A state $\ket{\psi}$ of length $L$ with periodic boundary conditions can then be constructed as follows:

\ie\label{eq:MPS_intro}
\vcenter{\hbox{\includegraphics[scale=0.25]{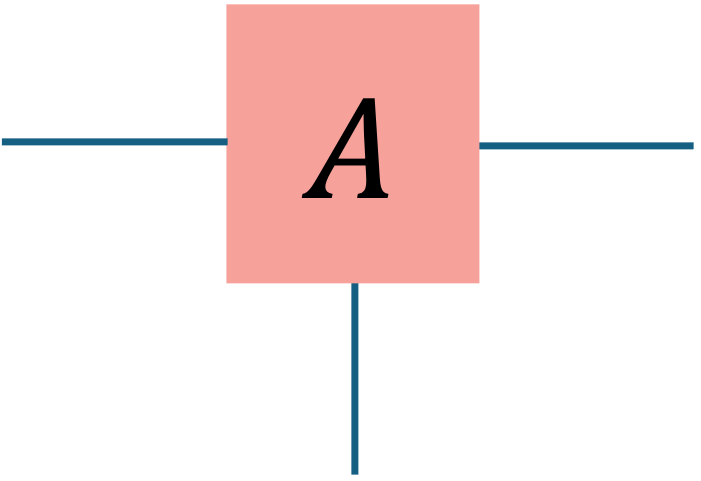}}} \ \implies \ \vcenter{\hbox{\includegraphics[scale=0.25]{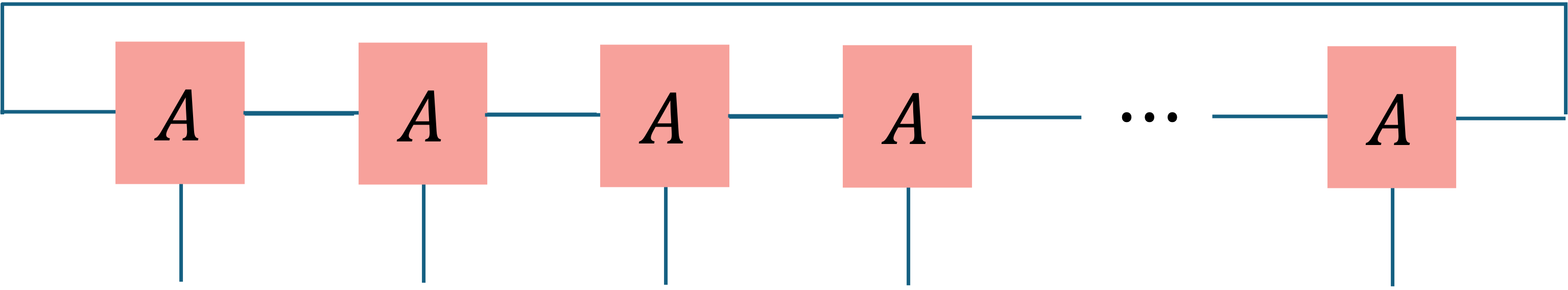}}} 
\fe

which represents the state
\begin{equation}
    \ket{\psi} = \sum_{s_1,\ldots,s_L} \Tr(A^{s_1}\ldots A^{s_L}) \ket{s_1\ldots s_L}.
\end{equation}

Multiple different tensors $A^{s}$ can correspond to the same physical periodic state. One usually works with the left-canonical form defined by the following conditions:

\ie\label{eq:canonical}
\vcenter{\hbox{\includegraphics[scale=0.25]{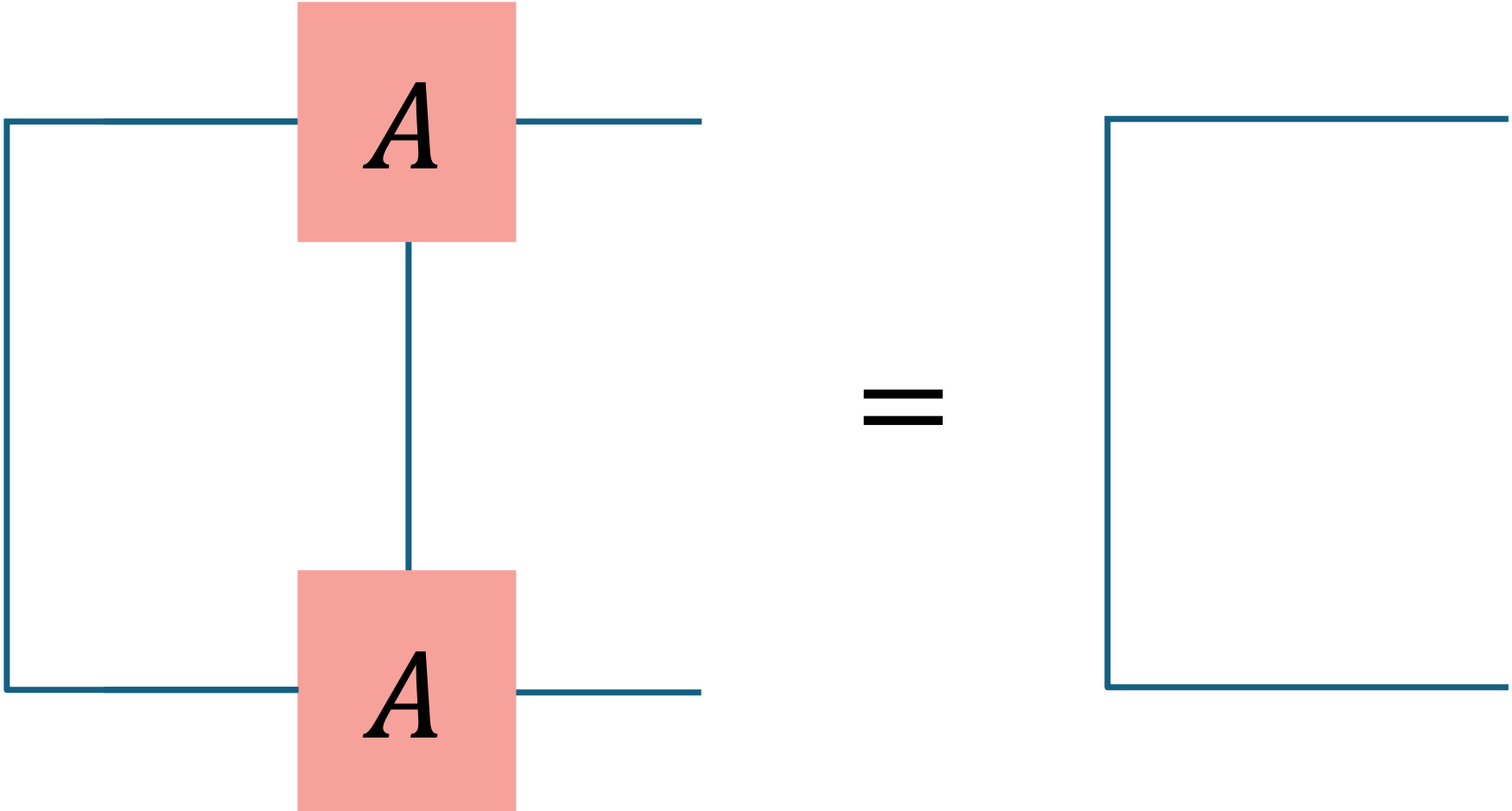}}}\ \,\,\,\, ,  \  \vcenter{\hbox{\includegraphics[scale=0.25]{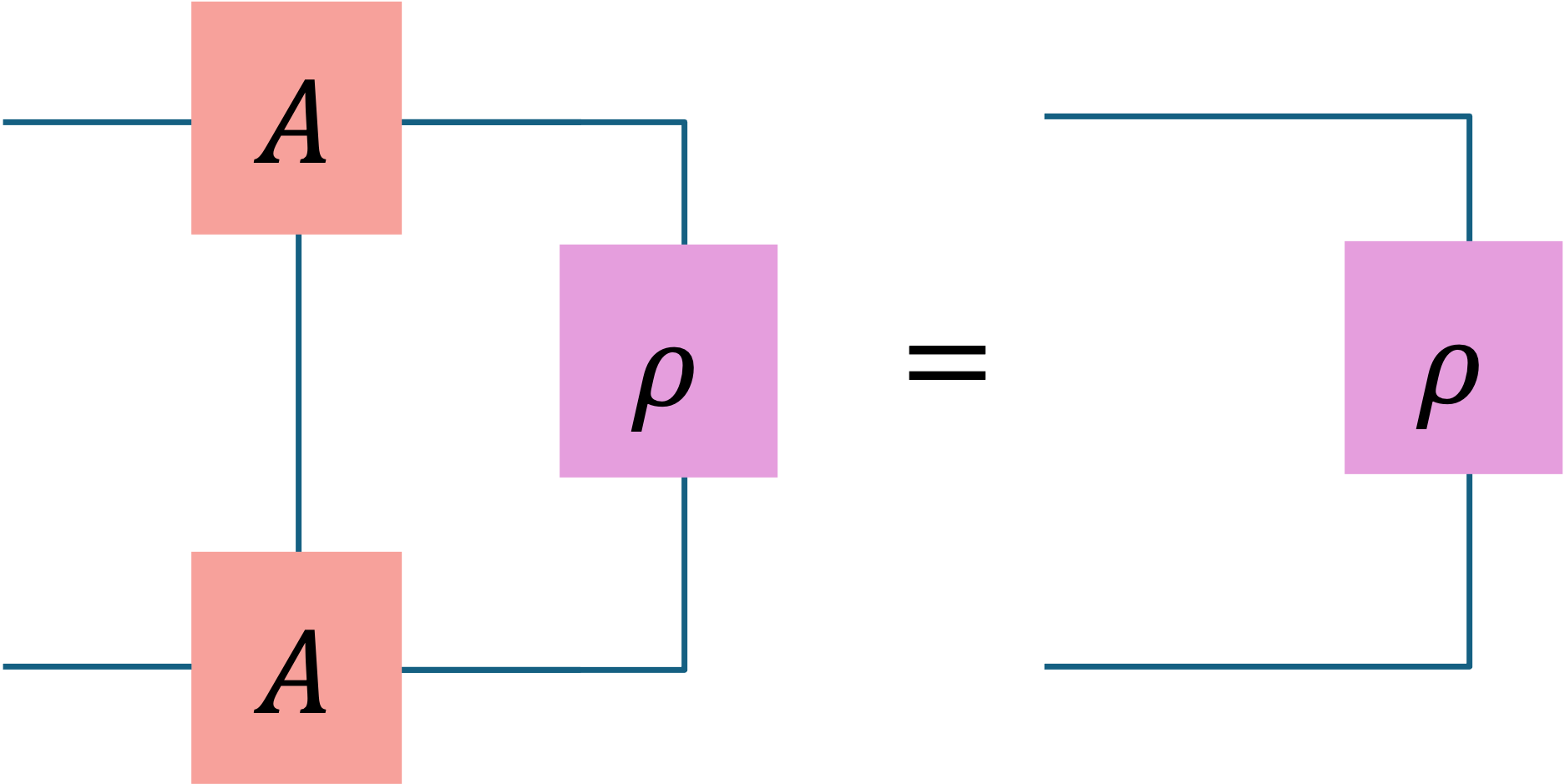}}}
\fe
which can be interpreted as an eigenvalue condition for a transfer matrix; these left and right eigenvectors of the transfer matrix are dominant eigenvectors. Any translationally invariant MPS can be transformed into this form without changing the physical state. The matrix $\rho$ is full-rank and positive definite. The MPS is called \textit{injective} if these dominant eigenvectors are unique (i.e. all other transfer matrix eigenvalues $\abs{\lambda}<1$). If a translationally-invariant injective MPS carries a (possibly modulated) unitary on-site symmetry\footnote{For modulated symmetries, it is a slightly subtle matter to describe how an MPS is symmetric, since modulated symmetries generally do not leave periodic MPS invariant. Nevertheless, with an appropriate definition of what it means for an MPS to be symmetric, one can derive the same pushthrough condition in full generality-- see \cite{anakru2026matrix}. For dipole/multipole symmetries in particular, there is no subtlety with periodic boundary conditions since the pushthrough result can be easily derived for open chains (\cite{ho_tat_lam_dipolar_SPT_classification,Saito_Multipolar_MPS}), which can always be taken to be exactly invariant under the unitary symmetry.}, then the on-site unitary `pushes through' the MPS tensor in the following way:

\ie\label{eq:Grand_Pushthrough_App}
\vcenter{\hbox{\includegraphics[scale=0.35]{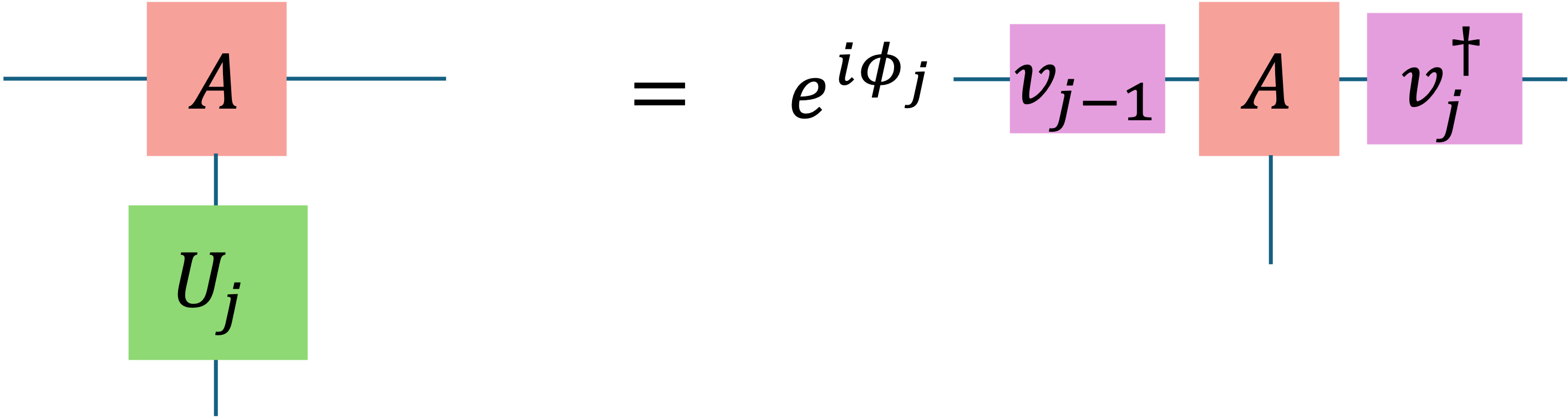}}}
\fe
with $v_j$ unitary. Unmodulated symmetries have $U_j$ independent of $j$, and thus $v_j$ is also independent of $j$. A dipole symmetry has $U_j = (U)^j$, where $U$ is some reference unitary, and dipole symmetry together with translation invariance implies an unmodulated charge symmetry generated by the on-site unitary $U$. One can show following arguments in \cite{ho_tat_lam_dipolar_SPT_classification} that we have the following pushthrough rules for a translationally-invariant injective MPS tensor with charge and dipole symmetry:

\ie\label{eq:charge_pushthrough}
\vcenter{\hbox{\includegraphics[scale=0.35]{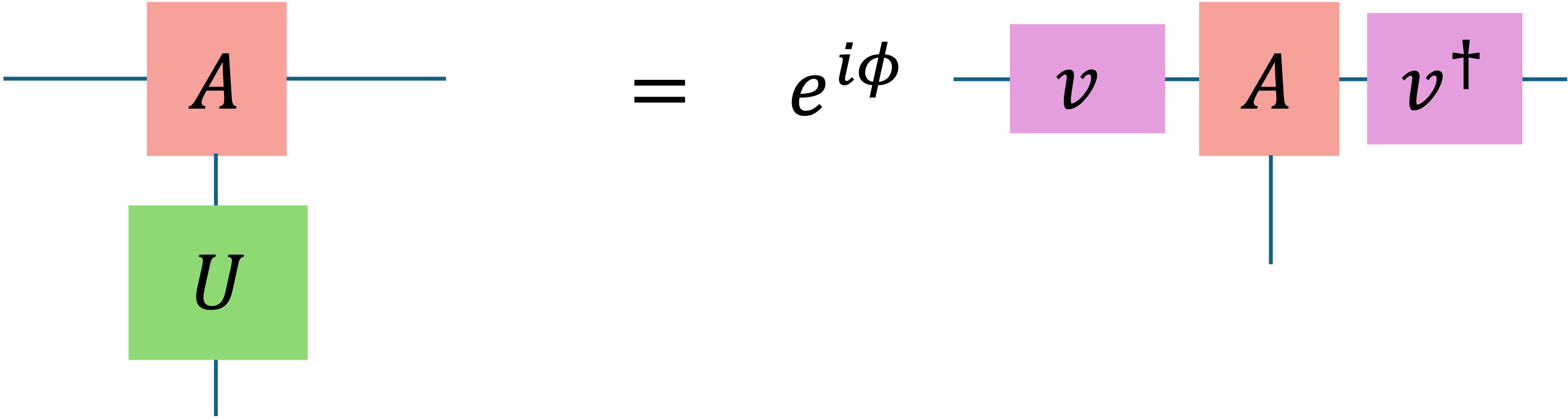}}}
\fe

\ie\label{eq:dipole_pushthrough}
\vcenter{\hbox{\includegraphics[scale=0.35]{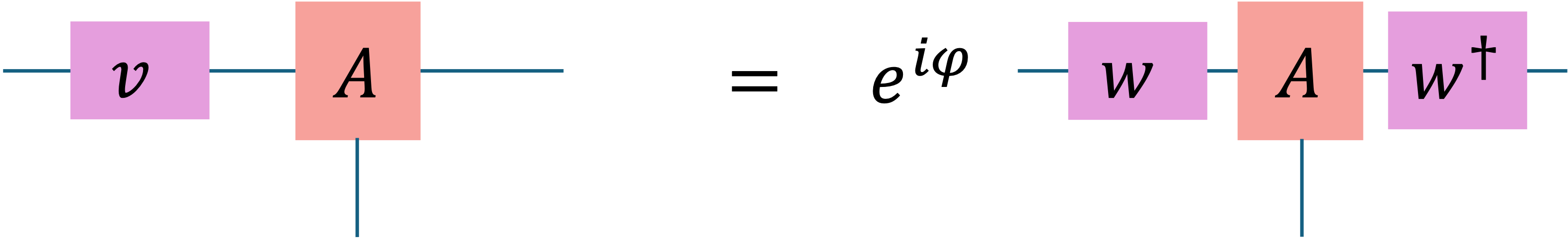}}}.
\fe
Here $\phi(\theta)$ and $\varphi(\theta)$ denote MPS
symmetry phases, distinct from the continuum fields used elsewhere.
Their derivatives at the identity,
$\nu_c=\partial_\theta\phi(0)$ and
$\nu_d=\partial_\theta\varphi(0)$, give the charge density and
bulk dipole-density contribution in the chosen virtual-charge convention,
respectively\footnote{These relations follow by differentiating the
pushthrough identities at $\theta=0$. For an interval of $L$ sites,
the dipole moment per site also contains the origin-dependent contribution
$L^{-1}\sum_{j=j_0}^{j_0+L-1}j\nu_c$ and a virtual-boundary charge
contribution, as described in \cite{zaletel_mong_pollmann_dipole_dmrg}.
The origin-dependent contribution vanishes at charge neutrality.}. Due to the area law \cite{hastings_area_law}, we expect that any trivial gapped ground state (which includes all 1-D gapped SPTs) can be well-approximated by a translationally-invariant injective MPS with all the same symmetries as the ground state wavefunction. As an example, one may arrive at SPT classification by classifying the injective MPS compatible with certain symmetry representations \cite{ChenGuWen2011,cirac_classifying,ho_tat_lam_dipolar_SPT_classification,Saito_Multipolar_MPS,anakru2026matrix,lam-ning-2026}. Also, some LSM theorems can be proved by assuming a symmetry representation on a translationally invariant injective MPS and finding a contradiction with injectivity and/or translation symmetry \cite{cirac_review,prakash2020elementary}.   \hfill \break

Here we will follow the spirit of these arguments by assuming that an injective MPS has charge ($U(1)_c$), dipole ($U(1)_d$), and particle-hole symmetry ($\mathbb{Z}_2^{\mc{P}}$). We also assume the existence of symmetry-protected fractionalized edge charges, which in the MPS language corresponds to bond unitaries $v(\theta)$ and $\sigma$ coming from $U(1)_c$ and $\mathbb{Z}_2^{\mc{P}}$ respectively, that transform in a nontrivial projective representation $v(\theta)\sigma = e^{i\theta}\times \sigma v(-\theta)$. Now we will see that using Eqs. \ref{eq:charge_pushthrough}-\ref{eq:dipole_pushthrough}, we can relate the edge projective representation to a bulk dipole filling, and arrive at a contradiction with injectivity. To see this, consider the following series of MPS manipulations:

\ie\label{eq:linhao_argument}
\vcenter{\hbox{\includegraphics[scale=0.35]{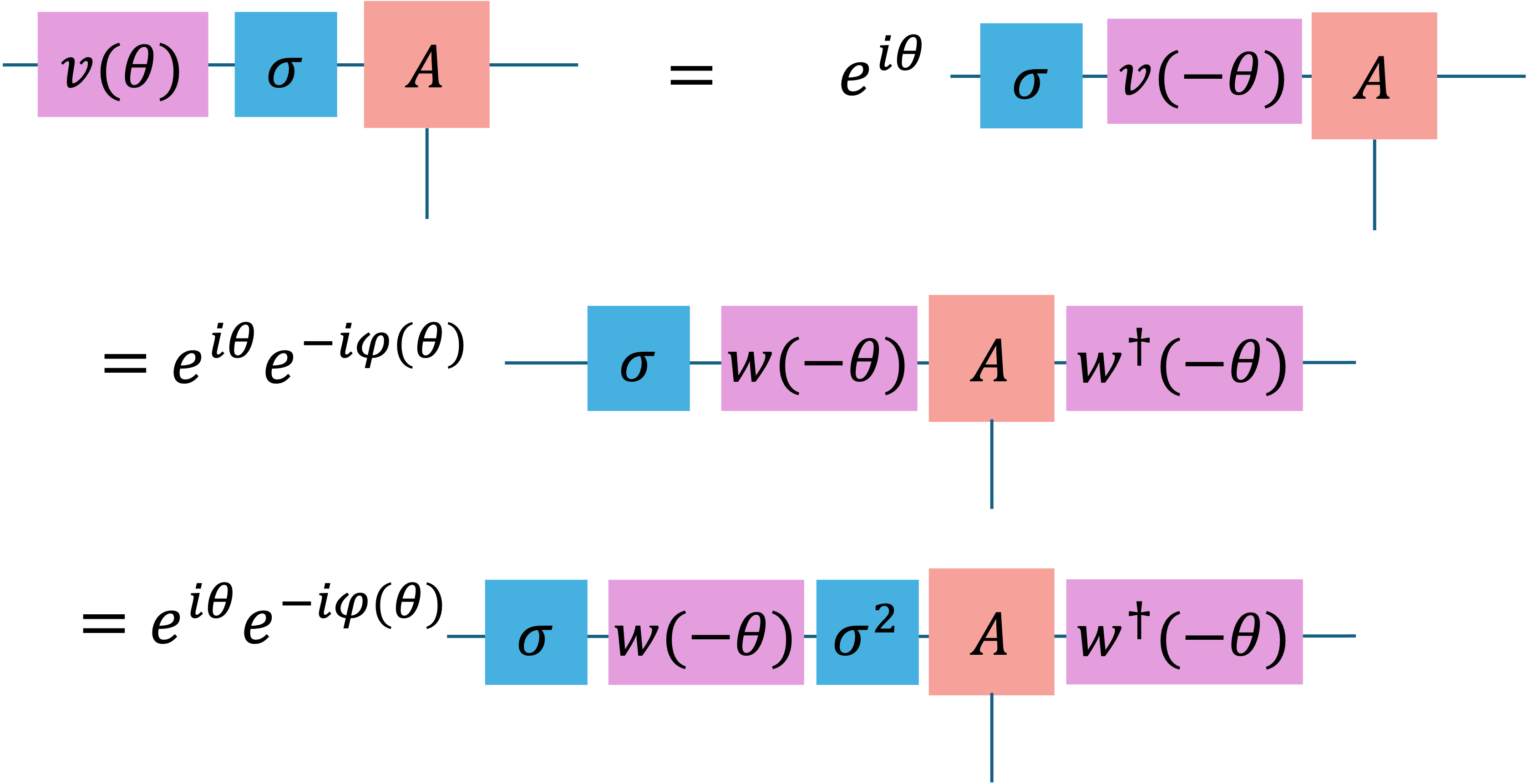}}} \\
\hfill
\vcenter{\hbox{\includegraphics[scale=0.35]{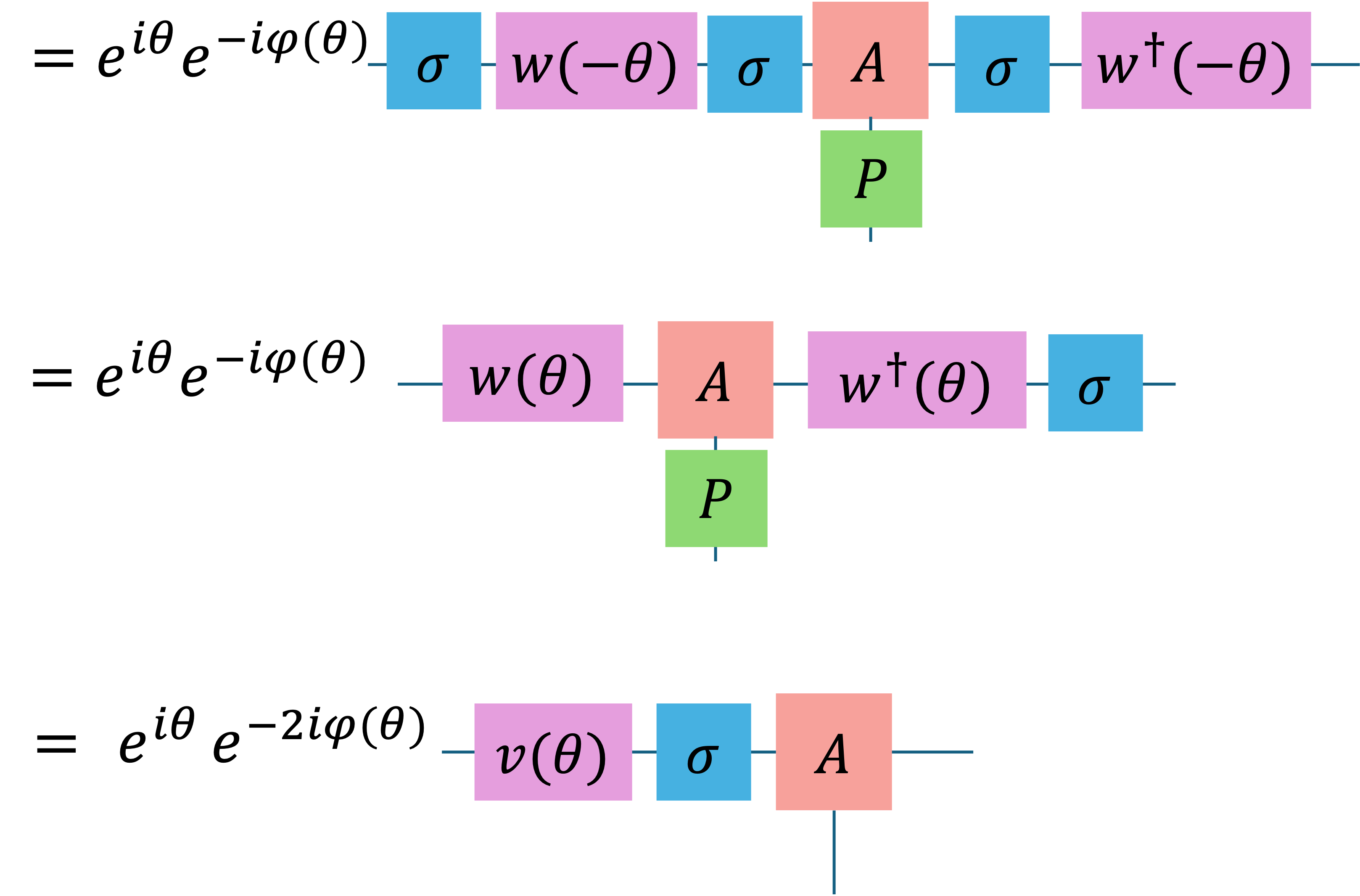}}}
\fe
where each step either uses the pushthrough relation for charge, dipole, or $\mathbb{Z}_2$ particle-hole, or the bond projective representation of the commutation relations in the symmetry algebra. The $e^{i\theta}$ that gets carried through is the assumed projective phase for the charge and particle-hole symmetries, which characterizes the putative gapped SPT phase (equivalently, the edge modes), and we have not assumed any projective phase between the dipole and particle-hole operators (it would anyway cancel out in the above manipulations). In the end, we are left with the statement that $e^{i\theta}=e^{2i\varphi(\theta)}$. In other words, a nontrivial projective representation of
$U(1)_c\rtimes\mathbb Z_2^{(\mc P)}$ implies
$\varphi(\theta)=\theta/2$ modulo $\pi$. Continuity and the identity
normalization fix the local branch near $\theta=0$, giving
$\nu_d=1/2$, which corresponds to a \textit{half-integer} dipole density\footnote{If we modified the projective representation furnished by $v$ by a 2-coboundary (which is the boundary of a 1-cochain), we can clearly see from Eq. \ref{eq:dipole_pushthrough} that the phase function $\varphi(\theta)$ shifts by the corresponding 1-cochain. This fully parametrizes the freedom in the projective representation, hence this relation between $\theta$ and $\varphi$ is robust.}. However, the phase factor $e^{i\varphi(\theta)}$ defined in Eq.~\ref{eq:dipole_pushthrough} must be $2\pi$-periodic in $\theta$, which is incompatible with the result $\varphi(\theta)=\theta/2$ obtained above.

One may also be tempted to interpret the above arguments as a proof that half-quantized edge charges (i.e. projective representation for $U(1)_c\rtimes \mathbb{Z}_2^{(\mc{P})}$) imply a bulk dipole density, as we have claimed in the main text. However, all of our above manipulations assumed injectivity, and the point of the argument was to rule out either injectivity or edge charges. There is a way to turn the above argument into a direct proof of the relationship between edge charges and bulk dipole density, but it requires further assumptions. If we assume an injective MPS that is only invariant under translation by two sites, along with the same pushthrough rules in Eqs. \ref{eq:charge_pushthrough}-\ref{eq:dipole_pushthrough} (with the bond matrices and phase functions modified to reflect the reduced translation symmetry, but carrying the same projective phases and charge densities as before), then there is no contradiction with a half-quantized dipole density, since our reduced translation symmetry evades the general restrictions set forth in \cite{anakru2026matrix,lam-ning-2026}. Indeed, the (roughly) 2-periodic MPS that DMRG finds are precisely examples of injective MPS that satisfy both the required dipole density and the charge projective representation. We will not, however, claim a full `tensor network proof' of the stated relationship, because we do not yet have a full classification of non-translationally-invariant injective MPS (or translation symmetry-breaking MPS) under modulated symmetries.

\section{Dipole Quantum Numbers and Details of MPS}

All our DMRG numerics conserve both the total charge and total dipole moment, following the prescription of \cite{zaletel_mong_pollmann_dipole_dmrg} which we review below. At the tensor network level, charge conservation is implemented in the usual way, and dipole conservation is implemented by defining an on-site quantum number that is weighted by the position along the chain. The two quantum numbers are thus

\begin{equation}\label{eq:qn_onsite_DMRG}
\begin{split}
\hat Q&=\sum_j\hat q_j,\qquad \hat q_j=\hat n_j^c+\hat n_j^d-1,\\
\hat D&=a\sum_j j\hat q_j,\qquad
\hat D/a-L\rho_0=\sum_j(j\hat q_j-\rho_0).
\end{split}
\end{equation}
where $n_j^c,n_j^d\in\{0,1\}$ are the occupancies of the $c,d$ fermions, and $\rho_0$ is the background dipole moment per unit cell in units of $a$. The implemented charge $\hat Q$ and background-subtracted dipole $\hat D/a-L\rho_0$ have associated bond quantum numbers $q_c^{(j)}$ and $q_d^{(j)}$, and the MPS tensor must conserve total charge and dipole. In a translationally invariant MPS, we would have the bond charges $q_c^{(j)}$ be independent of $j$, but the bond dipoles would obey $q_d^{(j+1)} = q_d^{(j)} + q_c$. For this reason, we often work with the `renormalized' bond quantum number $\tilde{q}_d^{(j)} =q_d^{(j)}-jq_c^{(j)}$, since in a uniform MPS it would be a truly translationally invariant object (this is also why we subtracted the average dipole density $\rho_0$ in Eq. \ref{eq:qn_onsite_DMRG}). 

We frequently plot entanglement spectra resolved by the renormalized bond dipole charge $\mathcal D_A=\hat D_A/a-j\hat Q_A-j\rho_0$ carried by the Schmidt eigenstates. This is easy to do with DMRG numerics because the eigenvalues $k$ of $\mathcal D_A$ carried by the Schmidt states are precisely the renormalized bond quantum numbers $\tilde q_d^{(j)}$. (One can see this by applying the MPS symmetry conditions Eqs. \ref{eq:charge_pushthrough}-\ref{eq:dipole_pushthrough}.) \hfb 

\begin{figure}[t]
  \centering
  \begin{subfigure}[t]{0.49\linewidth}
    \centering
    \includegraphics[width=\linewidth]{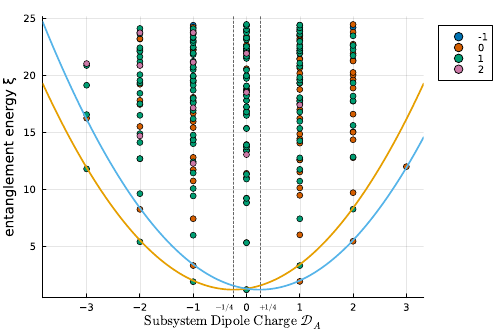}
    \caption{Even bond cut ($j=32$)}
    \label{fig:sm_even_bond_cut}
  \end{subfigure}\hfill
  \begin{subfigure}[t]{0.49\linewidth}
    \centering
    \includegraphics[width=\linewidth]{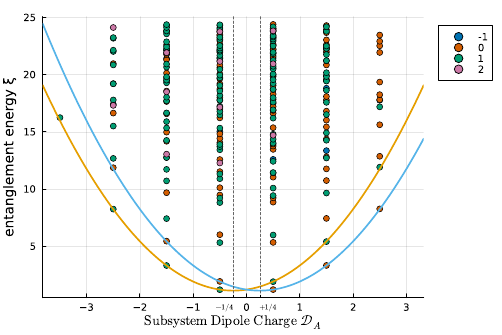}
    \caption{Odd bond cut ($j=31$)}
    \label{fig:sm_odd_bond_cut}
  \end{subfigure}
  \caption{Entanglement Spectrum of igSPT with parabolic fit to primary scaling dimensions, using $K\approx 1.47$. Color indicates the total ordinary $U(1)$ charge on the left half of the open chain, according to the legend. We take $h=t_d=1$, $\alpha = 1.3$, $U_0=U_1=-0.6$, $U'=1.2$, and $L=63$ with bond dimension $\chi=256$.}
  \label{fig:sm_even_odd_entanglement_spectra}
\end{figure}

We notice that for $\rho_0 = \frac12$, which is the case for the igSPT, dipole conservation forces the left and right bond dipole quantum numbers to differ by half-integers. This is not possible in an injective, translationally-invariant MPS\footnote{Indeed, this is the MPS manifestation of the LSM anomaly \cite{cirac_review}.}. In fact, a symmetric MPS regularization of the igSPT phase would need to spontaneously break translation symmetry $\mathbb{Z}\to 2\mathbb{Z}$; in practice, the DMRG computation picks out a wavefunction that explicitly breaks translation symmetry. This translation breaking is negligible in the thermodynamic limit, but it leaves an imprint on the MPS in the form of an even-odd effect on the entanglement spectrum, where odd bond entanglement cuts carry half-integer values of $\tilde{q}_d$ and even bonds carry integer $\tilde{q}_d$. The entanglement spectrum for the odd bonds can be fit to the same parabolic envelopes $E_n(\tilde{q}_d) = \frac{1}{2K}(\tilde{q}_d\pm\frac14)^2 + n$ that we found for the even bonds, with the only difference being that we plug in a half-integer for $\tilde{q}_d$ instead of an integer. In Fig. \ref{fig:sm_even_odd_entanglement_spectra}, we show the entanglement spectrum (and its parabolic fit) for both the even and odd bonds nearest to the midpoint of the chain. We fit $K$ independently for both plots, and find agreement in $K$ up to $10^{-2}$ in Fig. \ref{fig:sm_even_odd_entanglement_spectra}. In all other computations, we get the Luttinger parameter by fitting the \textit{even} bond entanglement spectrum. We also note that, for an odd bond, the half-integer charge $\tilde{q}_d$ is related to the winding number $w\in\mathbb{Z}$ in the BCFT by $\tilde{q}_d=w\pm\frac{1}{2}$, consistent with the BCFT energy spectrum $E\propto (w\pm\frac{1}{4})^2$.

Finally, we note that the twofold degeneracy of Schmidt eigenvalues with bond dipole charge $\neq 0$ is slightly lifted. This is a result of edge effects propagating to the bulk (see Fig. \ref{fig:charge_distribution} in the main text), since the choice of edge eigenstates breaks global particle-hole symmetry. We expect that these edge effects vanish in the thermodynamic limit, as they are exponentially suppressed deep in the bulk.  

\hfill \break

\section{Infinite MPS Entanglement Spectrum and Boundary conformal field theory}
In this appendix we predict a quantized quarter shift in the conformal towers of the entanglement spectrum of an infinite matrix product state (iMPS), based solely on symmetry constraints. According to Ref.~\cite{lauchli_MPS_BCFT}, the entanglement spectrum of an iMPS corresponds to the energy spectrum of an associated boundary conformal field theory (BCFT). In this correspondence, one boundary condition is determined by the entanglement cut, which we refer to as the \emph{entanglement boundary}, while the other arises from the effective perturbation induced by the finite bond dimension and is referred to as the \emph{physical boundary}.

Following the correspondence between topological edge states and the entanglement spectrum, we assume that the entanglement cut inherits the two fractionalized boundary-charge sectors of the igSPT \cite{LiHaldane2008,PollmannTurnerBergOshikawa2010,TurnerPollmannBerg2011}. Particle-hole symmetry $\mathcal{P}$ exchanges these sectors because their boundary charges are opposite. 
After rescaling time to set the bulk velocity to unity, the dipolar Luttinger liquid is described by
\begin{equation}\label{eq:Goldstone_LL_APP}
\mathcal{L}_{\mathrm{eff}}[\varphi]
=\frac{K}{2\pi}
\left[(\partial_t\varphi)^2-(\partial_x\varphi)^2\right]
+\frac{\theta}{2\pi a}\partial_t\varphi,
\qquad \theta=\pi.
\end{equation}
We consider boundary conditions that preserve the dipole $U(1)$ symmetry. These are Neumann conditions for $\varphi$, or equivalently Dirichlet conditions for its dual field, for which we use the compactification convention $\vartheta\sim\vartheta+1$. Denote the two entanglement-boundary values by $\vartheta_1$ and $\vartheta_2$.

The Berry-phase term assigns momentum $\pi/a$ modulo $2\pi/a$ at $\theta=\pi$ to the vortex operator $e^{2\pi i\vartheta}$, so a one-cell translation acts on the dual field as $\vartheta\to\vartheta+1/2$ \cite{Haldane_1D_LSM_Field_Theory_Thing_Vortex_Momentum}. %
We further assume that the entanglement spectrum is independent of the location of the cut. In the minimal two-boundary-state description, we implement this requirement by taking the set of entanglement boundary conditions to be closed under this half-period shift:
\begin{equation}
\{\vartheta_1,\vartheta_2\}
=
\{\vartheta_1+\tfrac12,\vartheta_2+\tfrac12\}
\pmod{1}.
\end{equation}
Thus $\vartheta_2=\vartheta_1+1/2$ modulo $1$. 

Particle-hole symmetry acts as $\vartheta\to-\vartheta$. By the topological boundary identification above, it exchanges the two distinct boundary conditions, requiring $\vartheta_2=-\vartheta_1$ modulo $1$. Combining the two constraints gives
\begin{equation}
\vartheta_2=\vartheta_1+\frac12=-\vartheta_1
\pmod{1},
\qquad
\{\vartheta_1,\vartheta_2\}
=\left\{-\frac14,\frac14\right\}
\pmod{1}.
\end{equation}

On the other hand,  the iMPS here respects the dipole U$(1)$ and particle-hole symmetry, which should be also preserved by the physical boundary condition. Thus we can obtain
\[
\vartheta_L=0\text{ or }\frac{1}{2} \text{ (mod 1) }.
\]

The corresponding BCFT energy spectrum is given by \cite{digiulio2023boundary,lauchli_BCFT,lauchli_MPS_BCFT}
\begin{equation}\label{eq:BCFT_NEUMANN}
E_n^{(s)}(w)
\propto\left[\frac{(w+\delta_s)^2}{2K}+n\right],
\qquad w\in\mathbb{Z},~ n\in\mathbb{Z}_{\geq0}.
\end{equation}
The zero-mode shifts are determined by the relative boundary values, $\delta_s=\vartheta_{R,s}-\vartheta_L$, with $\vartheta_{R,s}=\vartheta_s$, giving $\{\delta_1,\delta_2\}=\{-1/4,1/4\}$ modulo $1$. This implies a quarter shift in the entanglement spectrum.

\subsection{Convergence of Finite MPS Entanglement Spectrum to iMPS Entanglement Spectrum}

All of our computations use finite-chain DMRG, which is expected to approximate the thermodynamic limit in the limit of large system size and large bond dimension. As commented before, the dipole-symmetric open-chain eigenstates necessarily break particle-hole symmetry on the boundary; this is an additional finite-size effect which causes small deviations from the BCFT entanglement spectrum. 

\begin{figure}[t]
  \centering
  \begin{subfigure}[t]{0.49\linewidth}
    \centering
    \includegraphics[width=\linewidth]{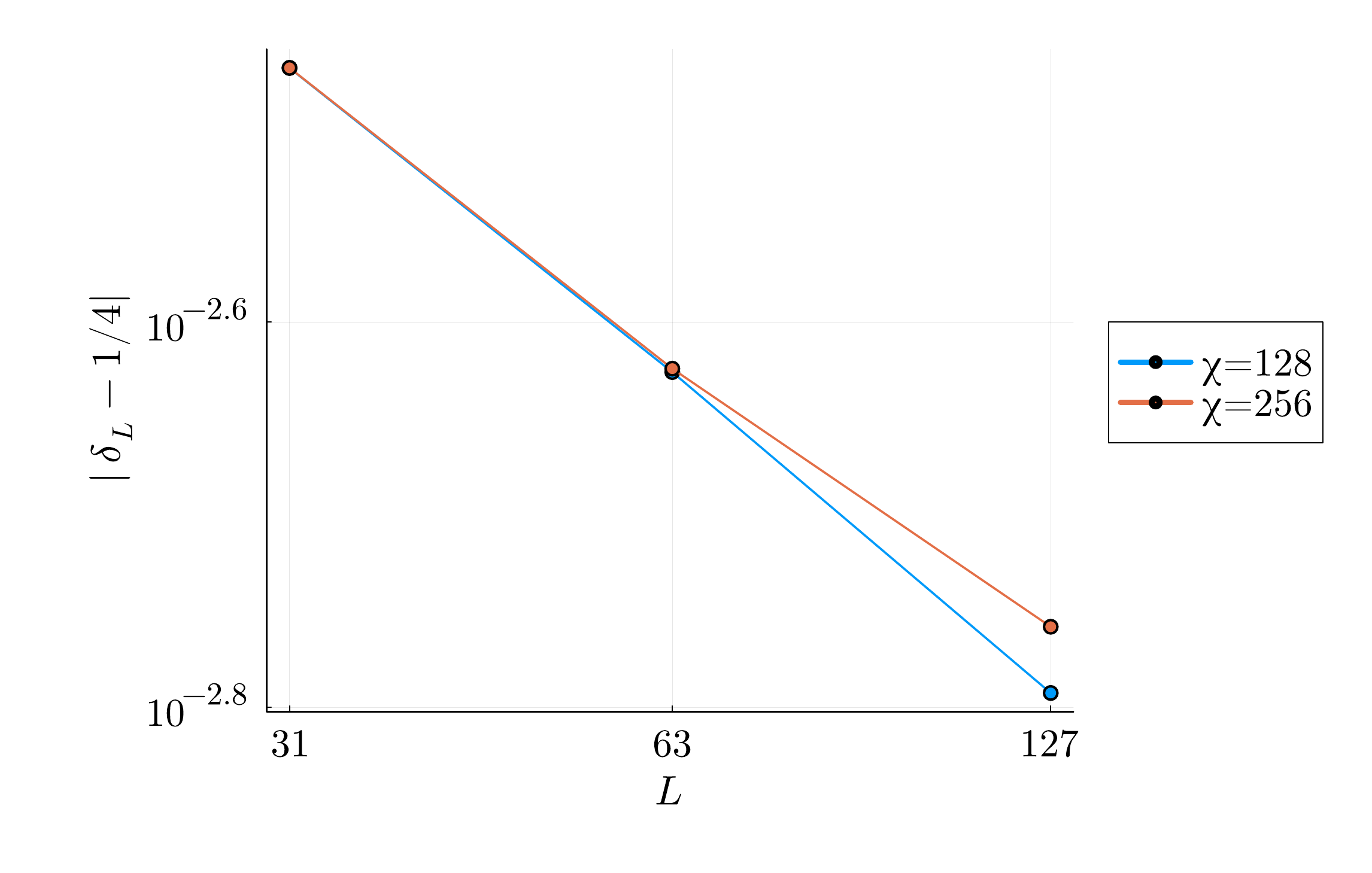}
    \caption{Left parabola shift $\delta_L$}
    \label{fig:delta_L}
  \end{subfigure}\hfill
  \begin{subfigure}[t]{0.49\linewidth}
    \centering
    \includegraphics[width=\linewidth]{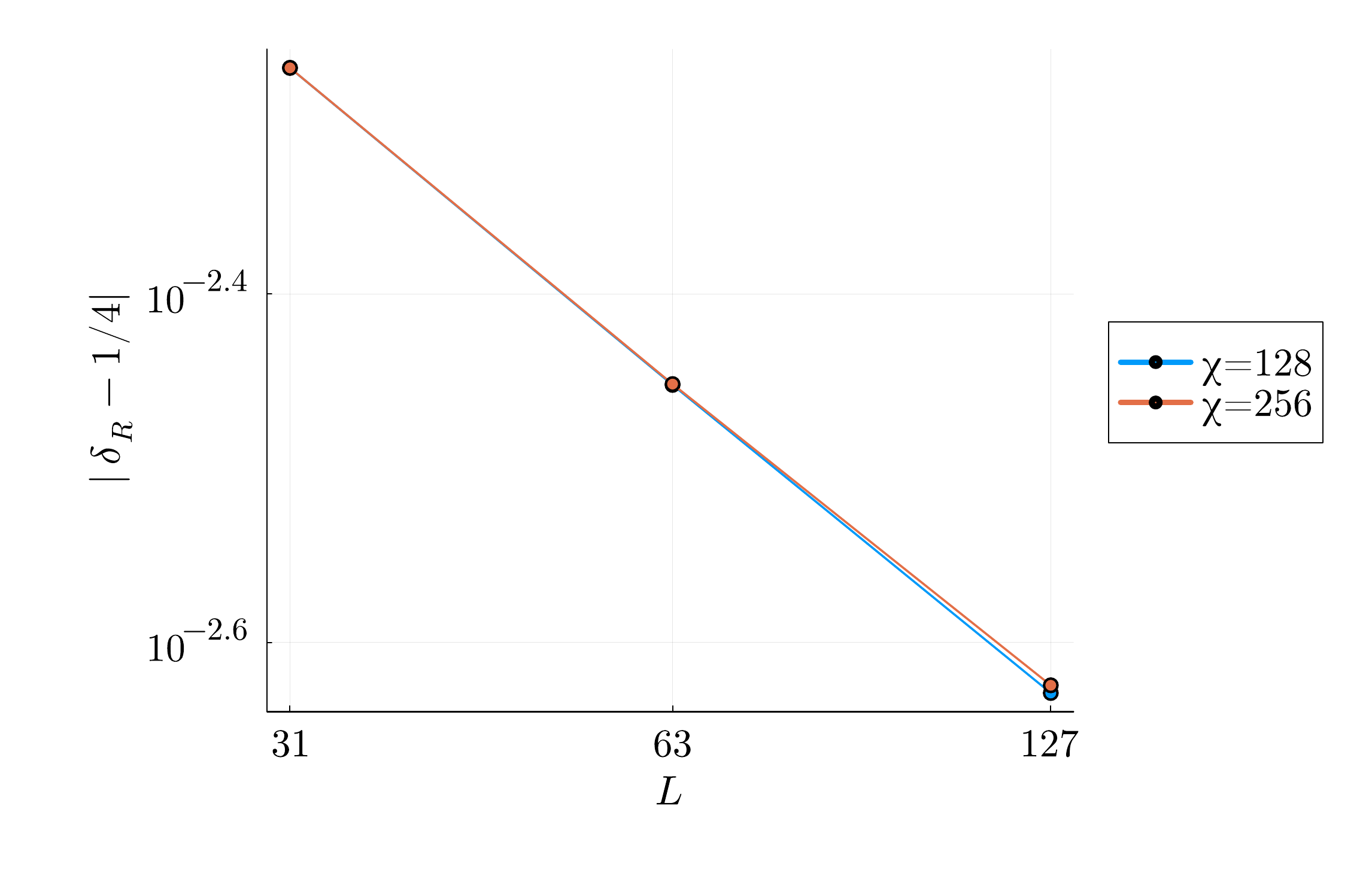}
    \caption{Right parabola shift $\delta_R$}
    \label{fig:delta_R}
  \end{subfigure}\hfill
  \caption{Fitted shifts $\delta_R,\delta_L$ of the BCFT spectrum, computed at the midpoint cut $j=(L+1)/2$ in all cases. We take $h=0.8$, $t_d=1$, $\alpha = 1$, $U_0=U_1=-0.6$, $U'=1.2$.}
  \label{fig:delta_scaling}
\end{figure}

Here, we show that such deviations are small and decrease with increasing system size $L$. We fit the two lowest-lying entanglement energies in each $\mathcal{D}_A$ sector to two parabolas $E_n(\tilde{q}_d) = \frac{1}{2K}(\tilde{q}_d +\delta_{L})^2, \frac{1}{2K}(\tilde{q}_d -\delta_{R})^2$ with two different shifts; here, $K$, $\delta_L$, and $\delta_R$ are all fitted parameters. In Fig. \ref{fig:delta_scaling}, we plot the deviation of the fitted values of $\delta_L,\delta_R$ from the BCFT prediction of $\frac14$, and observe that the deviations approach zero with increasing $L$. This indicates that our finite MPS calculations are indeed probing the same BCFT spectra obtainable from an iMPS calculation. 

Interestingly, the deviations in Fig. \ref{fig:delta_scaling} seem to decay as a power law. However, the DMRG at larger $L$ sometimes suffers from convergence issues since the system has gapless dipolar excitations -- even when the ground state energy is seemingly well-converged, the entanglement spectrum may be more strongly affected by mixing with such gapless excitations. Thus, we do not attempt to report a power law for the displayed scalings, only showing that the BCFT/iMPS predictions are indeed recovered as $L\to\infty$.

We note that the deviations at fixed $L$ at $\chi = 256$ are slightly larger than the deviations at $\chi = 128$; this can be simply explained by the fact that the entanglement spectrum has an increased density of states at higher bond dimension and thus has a higher susceptibility to particle-hole breaking perturbations. If we choose $\chi$ to scale with $L$ \cite{pollmann2009_finite_entanglement_scaling}, the increased susceptibility will only scale algebraically with $L$, whereas the effect of the edge perturbation on the midpoint entanglement spectrum decays \textit{exponentially} with $L$.

\section{Phase identification and further features of the phase diagram}\label{Sec:Phase_Diagram}

Here, we further discuss the phase diagram of Eq. \ref{eq:real_model_ssh_dipole} (with $\mathcal{A}_{ij} = \delta_{i,j+1} + \alpha\delta_{i,j+2} + h.c.$) as discussed in the End Matter. Alongside the results for the Luttinger parameter, we show the $q^3$-coefficient of the static structure factor, which we demonstrate is proportional to the Luttinger parameter for the topological part of the phase diagram (as per Eq. \ref{eq:charge_structure_factor}). When $\abs{Q_d}=\floor{L/2}$, we identify the gapped symmetry-breaking phase by the Luttinger parameter $K<0.5$. In the thermodynamic limit, one expects a KT transition at $K=0.5$ where double-vortices condense and destroy the dipolar quasi-long range order. At our finite size $L=31$, we expect that this identification procedure yields a good approximation to the phase boundary.

\begin{figure}[t]
  \centering
  \begin{subfigure}[t]{0.4\linewidth}
    \centering
    \includegraphics[width=\linewidth]{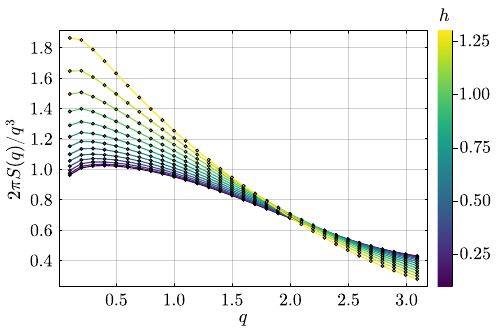}
    \caption{}
    \label{fig:supp_phase_diagram_DMRG_ssf_overlay}
  \end{subfigure}
    \hspace{0.04\linewidth}
  \begin{subfigure}[t]{0.4\linewidth}
    \centering
    \includegraphics[width=\linewidth]{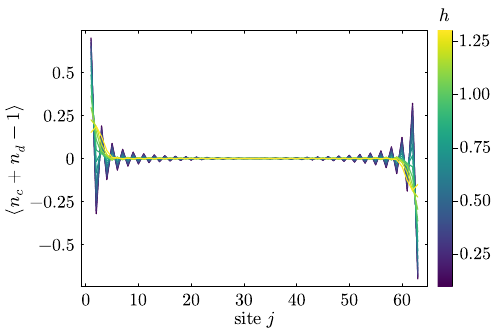}
    \caption{}
    \label{fig:supp_phase_diagram_DMRG_charge_density}
  \end{subfigure}

  \caption{
  Further observables of the model in Eq.~\ref{eq:real_model_ssh_dipole}
  with $t_d=1,U_0=U_1=-0.6,U'=1.2$.
  DMRG computations were done with $L=63$, $\chi=256$ and OBC.
  (a) DMRG result for the rescaled static structure factor $2\pi S(q)/q^3$ plotted for various values of $h$, at $\alpha=1$.
  (b) DMRG result for the charge density for various values of $h$, at $\alpha=1$.
  }
  \label{fig:supp_extra_observables}
\end{figure}

We also use exact diagonalization to verify the validity of our DMRG and to compute the energies of dipole symmetry sectors away from $\rho_d=0,\frac12$. In both DMRG and ED, we verify that sectors with dipole charges $\pm Q_d\neq 0$ are exactly degenerate (hence establishing the exact degeneracy of the edge modes at half-filling). Near the transition line, exact diagonalization on OBC suggests that a different dipole sector ($-\floor{L/2}+1$ as opposed to $\floor{L/2}$) may contain the true ground state; however, we attribute this to a finite size effect stemming from the edge modes (indeed, this effect is absent in PBC exact diagonalization, where edge effects are absent). Along with the Luttinger parameter growing as $h$ is tuned towards the phase boundary, the static structure factor seemingly crosses over to a scaling form $S(q)\sim q^2$ near the phase boundary (see Fig. \ref{fig:supp_phase_diagram_DMRG_ssf_overlay}).

\begin{figure}[t]
  \centering

  \begin{subfigure}[t]{0.35\linewidth}
    \centering
    \includegraphics[width=\linewidth]{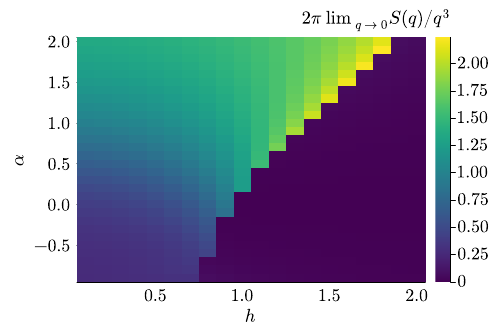}
    \caption{}
    \label{fig:supp_phase_diagram_ssf}
  \end{subfigure}
  \hspace{0.04\linewidth}
  \begin{subfigure}[t]{0.35\linewidth}
    \centering
    \includegraphics[width=\linewidth]{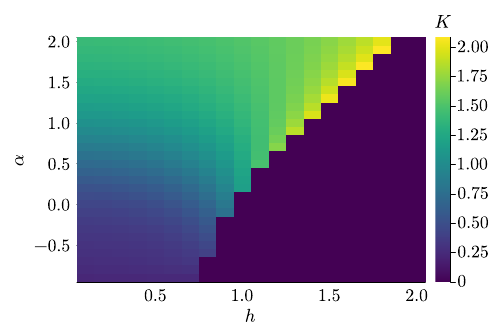}
    \caption{}
    \label{fig:supp_phase_diagram_K}
  \end{subfigure}

  \par\medskip

  \begin{subfigure}[t]{0.35\linewidth}
    \centering
    \includegraphics[width=\linewidth]{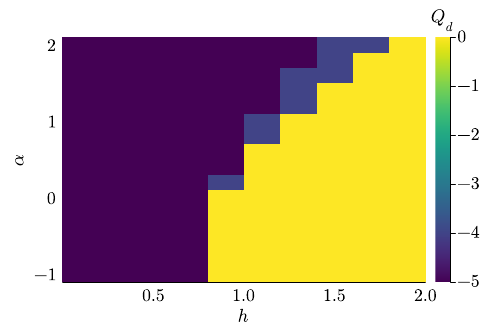}
    \caption{}
    \label{fig:supp_phase_diagram_ED_Q}
  \end{subfigure}
  \hspace{0.04\linewidth}
  \begin{subfigure}[t]{0.35\linewidth}
    \centering
    \includegraphics[width=\linewidth]{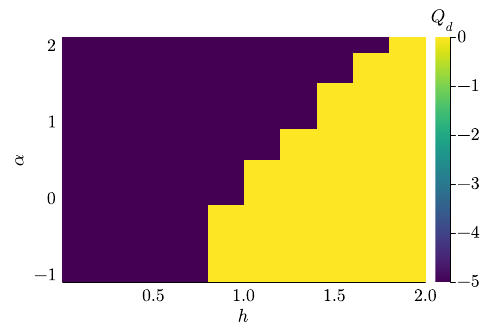}
    \caption{}
    \label{fig:supp_phase_diagram_ED_Q_PBC}
  \end{subfigure}
  \hspace{0.04\linewidth}
  \begin{subfigure}[t]{0.35\linewidth}
    \centering
    \includegraphics[width=\linewidth]{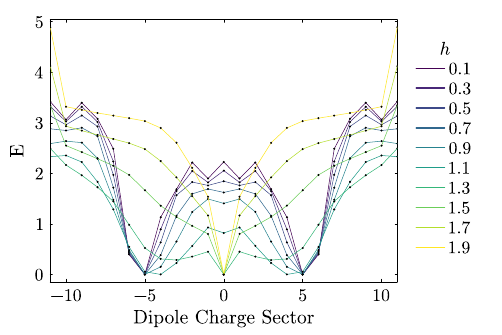}
    \caption{}
    \label{fig:supp_phase_diagram_ED_D_SPECTRUM}
  \end{subfigure}
    \hspace{0.04\linewidth}
  \begin{subfigure}[t]{0.35\linewidth}
    \centering
    \includegraphics[width=\linewidth]{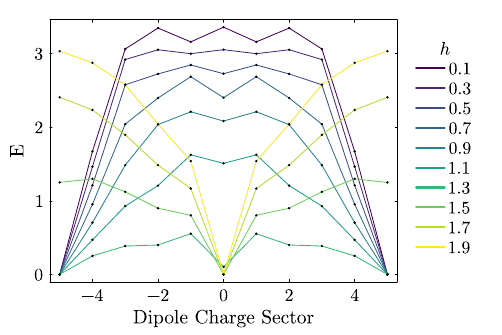}
    \caption{}
    \label{fig:supp_phase_diagram_ED_D_SPECTRUM_PBC}
  \end{subfigure}

  \caption{
  Various observables of the model in Eq.~\ref{eq:real_model_ssh_dipole}
  with $t_d=1,U_0=U_1=-0.6,U'=1.2$.
  DMRG computations were done with $L=31$, $\chi=128$ and OBC,
  and ED computations were done with $L=11$ and either OBC or PBC.
  (a) DMRG result for the $q^3$-coefficient of the static structure
  factor $S(q)$, scaled by $2\pi$. The coefficient is finite in the
  igSPT phase and vanishes in the thermodynamic limit in the trivial
  gapped phase.
  (b) DMRG result for the Luttinger parameter $K$, extracted from the
  mid-chain entanglement spectrum; within the given parameter range,
  this is close to panel~(a) up to finite-size corrections. In the trivial phase, we manually plot a Luttinger parameter of zero for visual distinction.
  (c) OBC ED result for the lowest-energy dipole symmetry sector. For the given size $L=11$, $Q_d=\pm5$ corresponds to the igSPT phase,
  whereas $Q_d=0$ corresponds to the trivial gapped phase.
  (d) PBC ED result for the lowest-energy dipole symmetry sector. Note that the dipole charge is now only defined modulo $L$.
  (e) OBC ED results for the lowest energy within each dipole charge
  sector, plotted for various values of $h$ with $\alpha=1$. 
  (f) PBC ED results for the lowest energy within each dipole charge
  sector, plotted for various values of $h$ with $\alpha=1$. 
  }
  \label{fig:supp_phase_diagram}
\end{figure}

\section{Scaling near the phase transition and possible Lifshitz criticality}
\label{Sec:1D_Ordered_Transitions}
Here, we propose a field-theoretical explanation for some of the behavior we saw in the previous section, i.e., the Luttinger parameter increasing near the phase boundary and the $q^2$-scaling behavior of $S(q)$ at intermediate momenta. While the numerical results suggest that the transition between the igSPT and trivial gapped phase is first order, we interpret this behavior as evidence for proximity to a critical point with $z=2$ scaling and one relevant tuning parameter. The understanding of the nature of this phase transition (and related ones) has appeared across \cite{Lake_Tilted_Chain,zechmann2022fractonic,anakru2023nonfermi,boesl2023deconfinement, jay_sau_double_well_boson,nahum2025_cts_ssb, andriyakhina2026theory}, and we summarize the relevant arguments here. 

We set $a=1$ in the following continuum derivation.
Taking Eq. \ref{eq:Mid_IR_ungauged} as a starting point, we treat both fermions and Goldstones as low-energy degrees of freedom and derive the following continuum Lagrangian, which resembles the action of a massive Dirac fermion coupled to a pseudo-gauge field:
\begin{equation}\label{eq:DiracAction}
\begin{split}
\mathcal{L}&=\bar{\psi}\gamma^\mu(i\partial_\mu-\varphi_\mu) \psi+\bar{\psi}me^{i\theta\gamma^5}\psi + \mathcal{L}_{\varphi}\\
\mathcal{L}_{\varphi} &= \frac12\kappa_0 (\partial_t\varphi)^2 -\frac12K_0(\partial_x\varphi)^2.
\end{split}
\end{equation}
Here $\psi=(\psi_{c},\psi_d)$, $\{\gamma^0,\gamma^1, \gamma^5\}=\{\sigma_y,-i\sigma_z,\sigma_x\}$, $\partial_\mu=(\partial_t,\partial_x)$, $\varphi_\mu = (0,-\varphi)$, and we have assumed an appropriate rescaling of the spatial coordinate. The angle $\theta\in\{0,\pi\}$ is the Berry phase of the filled fermion band.

Because the massive Dirac fermion is gapped, we may integrate it out by computing the fermion determinant and expressing it in terms of diagrams with external $\varphi$-legs, in the usual manner. The quadratic derivative terms in
Eq.~\eqref{eq:Quadratic_EFT_ungauged_Final} follow from the two-point
polarization of the free massive Dirac fermion, expanded at frequencies
and momenta small compared with $|m|$. In this approximation,
$\kappa-\kappa_0\propto m^{-2}$. The topological term is fixed separately
by the regulated fermion response, as made explicit below. At face value, this seems to reproduce our numerical observations, since this describes a Luttinger parameter that diverges near the critical point. It is shown in \cite{anakru2023nonfermi} that certain correlation functions are exactly obtainable when $m=0$, and they correspond to a Lifshitz theory with dynamical exponent $z=2$ (and the continuous dipole symmetry is spontaneously broken). However, what is not apparent in this fermionic language is that the Lifshitz critical point with $m=0$ is very unstable, with many relevant operators that drive the system back into the Luttinger liquid phase.

We also obtain the low-energy theory by bosonizing
Eq.~\eqref{eq:DiracAction} \cite{coleman_massive_schwinger}.
We normalize the charge boson by
$j^t=\partial_x\phi/\sqrt\pi$ and
$j^x=-\partial_t\phi/\sqrt\pi$, so a unit-charge winding has
$\Delta\phi=\sqrt\pi$. We use the corresponding compactification
$\phi\sim\phi+\sqrt\pi$, while the microscopic dipole phase retains
$\varphi\sim\varphi+2\pi$.
To keep track of the topological response, introduce
$\widetilde A_\mu=(A_t,\varphi)$ and
$\mathcal E_\varphi=\partial_t\varphi-\partial_xA_t$.
The current coupling satisfies
\[
\widetilde A_\mu j^\mu
=\frac{\phi}{\sqrt\pi}\mathcal E_\varphi
+\frac{1}{\sqrt\pi}
\left[\partial_x(A_t\phi)-\partial_t(\varphi\phi)\right].
\]
For the bulk topological response we use the regulated bosonized
coupling $\phi\mathcal E_\varphi/\sqrt\pi$, with the
$\theta=0$ band defining the trivial reference. Its normalization is
fixed by the fermion anomaly; the total derivative in the above
identity must be retained when matching boundary actions
\cite{coleman_massive_schwinger,qi_hughes_zhang_topological_field_theory}.
The bulk action is then
\begin{equation}\label{eq:DiracActionBosonized}
\begin{split}
\mathcal S=\frac12\int dt dx\biggl[
& (\partial_t\phi)^2-(\partial_x\phi)^2
+cm^2\cos(2\sqrt\pi\phi-\theta)
+\frac{2}{\sqrt\pi}\phi\mathcal E_\varphi+\frac{1}{g^2}\left((\partial_t\varphi)^2
-v_B^2(\partial_x\varphi)^2\right)\biggr].
\end{split}
\end{equation}
Here $g^{-2}=\kappa_0$ and $v_B^2/g^2=K_0$.
The normalization of the mass vertex, denoted by $cm^2$, depends
on the ultraviolet normal-ordering convention. Writing
$\phi=\phi'+\theta/(2\sqrt\pi)$ removes $\theta$ from the cosine
and gives
\[
\frac{\phi}{\sqrt\pi}\mathcal E_\varphi
=\frac{\phi'}{\sqrt\pi}\mathcal E_\varphi
+\frac{\theta}{2\pi}\mathcal E_\varphi.
\]
Below the charge gap, integrating out $\phi'$ generates higher-derivative
response terms, leaving the topological term
$\theta\mathcal E_\varphi/(2\pi)$. At $A_t=0$, this agrees with
Eq.~\eqref{eq:Quadratic_EFT_ungauged_Final} in units $a=1$.
Restoring $a$ gives
$\theta(\partial_t\varphi-a\partial_xA_t)/(2\pi a)$.
In the following, $\phi$ denotes fluctuations about the chosen minimum.

Now we see that at the $m=0$ point (which is Gaussian and a Lifshitz theory), there is not only an instability to the cosine term $\cos(2\sqrt{\pi}\phi)$ but there are further instabilities to arbitrary symmetry-allowed cosines $\cos(2p\sqrt{\pi}\phi)$ which could arise from fermion-fermion interactions. Since the field $\phi$ acquires a finite scaling dimension at the Gaussian Lifshitz point, we expand the periodic potential $V(\phi)=\sum_{\ell\geq1}\lambda_\ell
\cos(2\ell\sqrt\pi\phi)$. About the chosen minimum, absorbing numerical factors into the coefficients gives
$V(\phi)=V_0+\frac12(r\phi^2+s\phi^4+y\phi^6+\ldots)$ and treat these terms as perturbations to
Eq.~\eqref{eq:DiracActionBosonized}.
At zero background gauge field, Wick rotation $t=-i\tau$ gives the
quadratic Euclidean action
\[
S_{E,0}=\frac12\int d\tau dx\left[
(\partial_\tau\phi)^2+(\partial_x\phi)^2
+\frac{(\partial_\tau\varphi)^2+v_B^2(\partial_x\varphi)^2}{g^2}
+\frac{2i}{\sqrt\pi}\varphi\partial_\tau\phi\right],
\]
where we have integrated the mixed term by parts for the local bulk
fluctuations. Integrating out $\varphi$ yields
\[
S_{E,0}^{\mathrm{eff}}
=\frac12\int\frac{d\omega_E\,dp}{(2\pi)^2}
\left[\omega_E^2+p^2+
\frac{g^2\omega_E^2}{\pi(\omega_E^2+v_B^2p^2)}\right]
|\phi(p,\omega_E)|^2.
\]
For $|\omega_E|\ll v_B|p|$, the nonlocal temporal term dominates
$\omega_E^2$ at small momentum. Including the potential gives
\begin{equation}\label{eq:DiracActionBosonized_EFF_WF}
\begin{split}
S_E=\frac12\int d\tau dx\biggl[
&\frac{g^2}{\pi v_B^2}
(\partial_\tau\partial_x^{-1}\phi)^2
+(\partial_x\phi)^2+r\phi^2+s\phi^4+y\phi^6+\ldots\biggr].
\end{split}
\end{equation}
Here $\partial_x^{-1}$ is defined on nonzero spatial Fourier modes;
the zero mode is treated separately. It is easy to see that at the Gaussian point of Eq. \ref{eq:DiracActionBosonized_EFF_WF} with $r=0$, using the fact that $\partial_x\phi$ is proportional to the microscopic charge density, the static structure factor $S(q)\sim q^2$. Now the question is whether there is another Lifshitz theory that is more stable than Eq. \ref{eq:DiracActionBosonized_EFF_WF} (i.e. less relevant perturbations), such that Eq. \ref{eq:DiracActionBosonized_EFF_WF} can flow to such a theory when perturbed. To search for such a theory, we perform a Wilsonian RG where we integrate out modes with $\abs{p}\in [\Lambda/b,\Lambda]$ over all frequencies. Under the na\"ive power counting at the Gaussian Lifshitz point where $[\tau]=-2$, $[x]=-1$, we have $[r]=2$, $[s]=1$, and $[y]=0$ with all higher powers of $\phi$ being irrelevant.  Anticipating that $r$ remains relevant and $s$ flows to a fixed point, we work in $2-\epsilon$ spatial dimensions (i.e. within an $\epsilon$-expansion about the upper critical dimension). The one-loop RG flow is given in \cite{jay_sau_double_well_boson,nahum2025_cts_ssb} and features a stable fixed point at ${s},{r}=O(\epsilon)$ and ${y} = 0$. Note that the unstable "Gaussian" fixed point corresponding to ${r}={s}={y}=0$ is described by Eq. \ref{eq:DiracActionBosonized}. To order $O(\epsilon)$, the dynamical exponent is unchanged from $z=2$ at the interacting fixed point, and the dipolar order parameter is still long-range ordered.

It is important to contrast this fixed point with the quantum Lifshitz model as described in \cite{gorantla_lam_seiberg_shao_dipole_lifshitz_tensor_fractons, LakeDBHM,Lake_Tilted_Chain,zechmann2022fractonic}. The quantum Lifshitz form follows by dualizing the full coupled
quadratic action with the cosine terms set to zero. In the real-time
derivative-coupling convention, the canonical momentum of $\phi$ is
$\Pi_\phi=\partial_t\phi-\varphi/\sqrt\pi$.
Introduce $\vartheta_{\mathrm c}$ through
$\Pi_\phi=\partial_x\vartheta_{\mathrm c}$ and its conjugate momentum
$\Pi_{\vartheta_{\mathrm c}}=\partial_x\phi$. The dual bulk Lagrangian is
\[
\mathcal L_{\mathrm{dual}}
=\frac12(\partial_t\vartheta_{\mathrm c})^2
-\frac12\left(\partial_x\vartheta_{\mathrm c}
+\frac{\varphi}{\sqrt\pi}\right)^2
+\frac{1}{2g^2}\left[(\partial_t\varphi)^2
-v_B^2(\partial_x\varphi)^2\right].
\]
The combination
$\varphi+\sqrt\pi\partial_x\vartheta_{\mathrm c}$ is massive.
Integrating it out and retaining the leading infrared terms gives
\[
\mathcal L_{\mathrm{QL}}
=\frac12(\partial_t\vartheta_{\mathrm c})^2
-\frac{\pi v_B^2}{2g^2}
(\partial_x^2\vartheta_{\mathrm c})^2+\ldots,
\]
which has quadratic dispersion. This Gaussian fixed point is unstable
to the perturbations $\cos(2p\sqrt\pi\phi)$. It was noted in \cite{gorantla_lam_seiberg_shao_dipole_lifshitz_tensor_fractons} that when the theory is realized on a finite size system, the ground state degeneracy scales with the system size. This is due to an emergent "dipole winding" symmetry that has a nontrivial commutation relation with the microscopic dipole symmetry. In a particular representation of the Lifshitz theory:
\begin{equation}\label{eq:DiracActionBosonized_EFF_NICER}
\begin{split}
S_E=\frac12\int d\tau dx\biggl[
&\frac{g^2}{\pi v_B^2}(\partial_\tau\tilde\phi)^2
+(\partial_x^2\tilde\phi)^2+r(\partial_x\tilde\phi)^2+s(\partial_x\tilde\phi)^4
+y(\partial_x\tilde\phi)^6+\ldots\biggr].
\end{split}
\end{equation}
which is a redefinition of Eq. \ref{eq:DiracActionBosonized_EFF_WF} with $\partial_x\tilde{\phi}=\phi$, the dipole winding symmetry is given by $\tilde{\phi}\to \tilde{\phi}+cx$ and only holds when $r,s,y,\ldots = 0$. Thus, the dipole winding symmetry and hence the system size-dependent GSD is only a feature of the Gaussian Lifshitz theory and not a feature of the non-Gaussian fixed point theory that we describe. 

The field theory of Eq. \ref{eq:DiracActionBosonized_EFF_NICER} has actually been studied before, with $\tilde{\phi}$ as the phase field of microscopic bosons with only $U(1)$ symmetry \cite{jay_sau_double_well_boson} and an unconventional "double-well dispersion" (for $r<0$). In the language of \cite{jay_sau_double_well_boson}, the domain walls of the $\tilde{\phi}$-field obey an \textit{emergent} dipole-conservation law, because the total dipole moment of the domain walls is mapped to the total momentum of the $\tilde{\phi}$-field. In our case, the $\tilde{\phi}$ field is a nonlocal object, but $\pi^{-1/2}\partial_x^2\tilde{\phi}$ is identified with the local fermion operator $\bar{\psi}\gamma^0\psi$, which is the density of the microscopic charges that are subject to an \textit{exact} dipole symmetry.

\end{document}